\documentclass[prd,aps,preprint,showpacs,nofootinbib,superscriptaddress,tightenlines]{revtex4-1}
\usepackage{amssymb,amsmath}
\usepackage{hyperref}
\usepackage[capitalise]{cleveref}
\usepackage{graphicx,subfigure}
\crefname{section}{Sec.}{Secs.}
\begin{document}
\title{\textbf{Parton energy loss in high energy hard forward processes in proton-nucleus collisions}}
\author{Tseh Liou}
\author{A.~H.~Mueller}
\affiliation{Department of Physics, Columbia University, New York, New York 10027, USA}

\begin{abstract}
We calculate the spectrum of energy loss differences due to gluon radiation in high energy hard forward processes in proton-nucleus collisions as compared to proton-proton collisions. We find that the nuclear induced energy loss scales linearly with the beam energy. We evaluate the spectrum and ``typical'' energy losses in a logarithmic and large $N_{c}$ approximation. The energy losses found appear large enough to be phenomenologically important.
\end{abstract}

\maketitle

\section{Introduction}
The topic of this paper is partonic energy loss in high energy hard collisions. The type of process we have in mind is exemplified by forward jet production in proton- (or deuterium-) nucleus collisions. For example, in comparing proton-nucleus collisions with proton-proton collisions to extract information on high energy nuclear shadowing \cite{Tuchin,Arsene} it is important to understand, as emphasized by Frankfurt and Strikman \cite{Frankfurt}, how much additional energy loss leading partons suffer in proton-nucleus collisions as compared with proton-proton collisions. Also, recently Arleo and Peign\'e \cite{Peigne,Arleo,Sami} (see also Ref.~\cite{Kolevatov}) have done a detailed phenomenological analysis of forward $J/\psi$ suppression at fixed target energies and suggest that the key to their successful description is additional energy loss of the leading gluon (which converts to a charm-anticharm pair eventually becoming the $J/\psi$) in nuclear targets as compared to proton targets. The issues surrounding energy loss in these examples are not closely related to the energy loss of jets produced in high energy heavy ion collisions \cite{Zakharov,BDMPS}. In the heavy ion case the jets are produced as bare quanta in a QCD medium and the energy loss is limited to a factor times $\hat{q}L^{2}$ where $\hat{q}$ is the transport coefficient and $L$ is the length of (hot) QCD matter that the jets transverse. In forward jet or quarkonium production a dressed parton approaches the target and it is the additional energy lost by this dressed parton due to the nuclear target which is at issue.

Recently, Arleo and Peign\'e have given convincing first principle arguments that the additional energy loss caused by nuclear targets in forward jet or quarkonium production should scale linearly with the energy of the beam. Energy loss growing linearly with beam energy has also been advocated in other analyses of these processes \cite{Frankfurt,Kopeliovich}, however in Ref.~\cite{Arleo} for the first time a first-principles calculation of the exact form and magnitude of the spectrum of energy losses induced by nuclear targets has been given.

The process considered in Ref.~\cite{Arleo} was very high energy $J/\psi$ production. The picture is that a gluon coming from the proton projectile converts into a charm-anticharm pair, either well before or well after reaching the nucleus. The charm-anticharm pair ultimately becomes a quarkonium state, but the essential calculation is that of the additional energy lost by the gluon or the charm-anticharm pair due to having a nuclear as opposed to a proton target. In fact the conversion of the gluon into the charm-anticharm pair was not dealt with in Ref.~\cite{Arleo}. Rather the hard process, where the gluon converts into a charm-anticharm pair, was simulated by a hard scattering of the incoming gluon, at the scale of the charm mass, and then a calculation was given for the difference, between nuclear and proton targets, of the probability that an additional softer gluon be radiated. 

In \cref{sec:qqjet} we revisit the gluon $\rightarrow$ quark-antiquark process, with either light-quark jets or heavy-quarks being produced. The nucleus is treated in a McLerran-Venugopalan approximation \cite{Gelis}. While the details of our calculation look quite different from those of Arleo and Peign\'e, our result given in \cref{eq:radspec,eq:radspecqs} exactly reproduces their result. At the end of this section we indicate that small-$x$ evolution can trivially be added to extend the McLerran-Venugopalan model with the saturation momenta of the nucleus $Q_{s}$ and the proton $Q_{s}(\textrm{P})$ being the only parameters that enter the calculation.

In \cref{sec:qgjet} we extend our calculation from gluon $\rightarrow$ quark-antiquark to quark $\rightarrow$ quark-gluon jet production. The calculational procedure follows that of \cref{sec:qqjet} and the result is given in \cref{eq:radspecqgone,eq:radspecqgtwo}. The result here is very close to our calculation in \cref{sec:qqjet} with the $2\alpha_{s}N_{c}/\pi$ in \cref{eq:radspec} being replaced by $8\alpha_{s}N_{c}/5\pi$ in \cref{eq:radspecqgone}. This is, perhaps, surprising as there appears to be no simple relationship between the color charges of the initial and final partons and the prefactors of our results. The reason why there is no simple rule is explained at the end of \cref{sec:qgjet}. Using the quark $\rightarrow$ quark-gluon processes as an example we also estimate the \textit{typical} additional energy loss, $\bar{\omega}$, with a nuclear as opposed to a proton target and find $\bar{\omega}_{\textrm{RHIC}}\simeq E/15$ and $\bar{\omega}_{\textrm{LHC}}\simeq E/25$ where $E$ is the initiating quark's energy. We note that our RHIC estimate is very similar to that obtained by Frankfurt and Strikman \cite{Frankfurt}. 

In \cref{sec:quarkonium} we briefly discuss energy loss in quarkonium production in the color evaporation model \cite{Fritzsch,Halzen,Gluck,Barger}. We have very little to say here since our technical result is the same as that of Ref.~\cite{Arleo} and they have already done the detailed phenomenology.

In \cref{sec:colorsinglet} we consider nuclear induced energy loss in a color singlet model of quarkonium production. As originally emphasized in Ref.~\cite{Kharzeev} (see also Ref.~\cite{Dominguez}) in production on nuclei it is relatively easy to get the three or more gluons (one from the projectile and two or more from the nuclear target) interacting with the charm-anticharm pair so that it can have the quantum numbers of the $J/\psi$. However, this is not the case for a proton target. Thus in our discussion of singlet model production of quarkonium it is not natural to discuss nuclear minus proton target differences. What we calculate is the energy loss spectrum $\omega dI/d\omega$ for $J/\psi$ production on a nucleus. We separate two cases. In \cref{eq:radspeccolorsingletone} we give the spectrum in case $Q_{s}^{2}/M^{2}$, with $M$ the charm mass, is less than 1 while in \cref{eq:radspeccolorsinglettwo} we give the spectrum when $Q_{s}^{2}/M^{2}\simeq 1$. We estimate $\bar{\omega}\simeq E/6$ in both cases, and it should be remembered that now we refer to absolute energy losses not differences between nuclear and proton targets.

Finally, some general comments on the limitations of our calculation. To begin, our calculation is made in a logarithmic approximation. That is values of $\omega$ in $\omega dI/d\omega$ must be small enough that there is a significant logarithmic integration in the transverse momentum (coordinate) of the radiated gluon. In the end we find typical values of $\omega/E$ small enough that this logarithmic approximation is not unreasonable, but like all leading logarithmic calculations it cannot be expected to be very reliable. For simplicity, we have only considered the case where the produced jets have the same longitudinal momentum. We do not expect large changes so long as their longitudinal momenta are comparable. Also, in some respects our discussion of energy loss is complementary to recent discussions of transverse momentum broadening effects in similar processes \cite{Xiao,Yuan,Marquet,Qiu,Berger,Kang}.

\section{Energy loss in gluon $\rightarrow$ quark-antiquark jet production in proton-nucleus collisions\label{sec:qqjet}}
In this section we shall evaluate the energy loss in gluon $\rightarrow$ quark-antiquark jet production in proton-nucleus collisions. To simplify the calculation we shall use the large $N_{c}$ approximation as the multiple scattering with the nucleus appears difficult to do beyond the large $N_{c}$ limit. Our focus is on forward two-jet production at the LHC where the transverse momentum of each jet, $M$, is much greater than the saturation momentum $Q_{s}$. We shall begin using a simple McLerran-Venugopalan model \cite{Gelis} for multiple scattering with the nucleus, but at the end we shall see that small-$x$ evolution is very straightforward to implement as our results will be completely expressed in terms of dipole $S$ matrices. The calculations done in this section easily extend from the light-quark jets which we do here to heavy-quark pair production and thus to the energy loss in onium production in proton-nucleus as compared to proton-proton collisions. This then is the basic ingredient in testing the effects of energy loss in a color evaporation picture of onium production along the lines pioneered by Arleo and Peign\'e \cite{Peigne,Arleo}.

In addition to the interactions producing the two jets we include an additional radiative gluon, the source of the energy loss. We shall in turn evaluate initial state radiation in both the amplitude and the complex conjugate amplitude, final state radiation in the amplitude and complex conjugate amplitude, and interference terms having initial state radiation in the amplitude (complex conjugate amplitude) and final state radiation in the complex conjugate amplitude (amplitude). 

\subsection{Purely initial state radiation}
\begin{figure}[h]
  \centering
  \subfigure[]{\label{fig:a}
  \includegraphics[width=12cm]{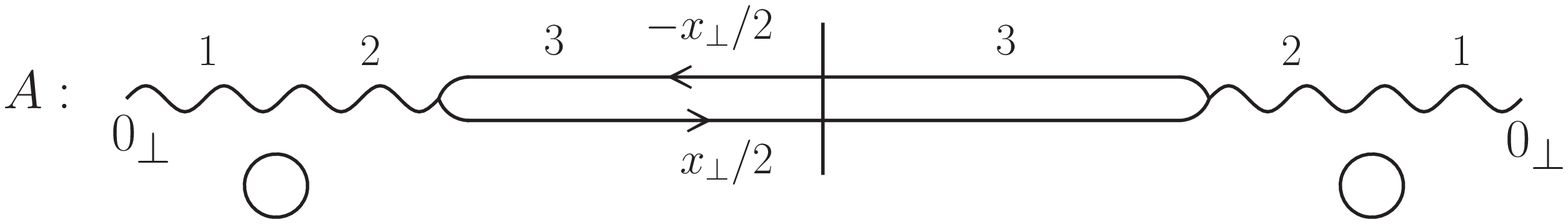}}
  \subfigure[]{\label{fig:b}
  \includegraphics[width=12cm]{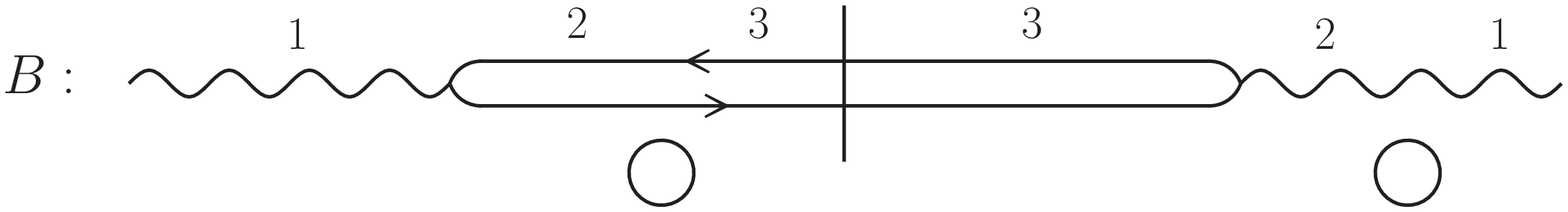}}
  \subfigure[]{\label{fig:c}
  \includegraphics[width=12cm]{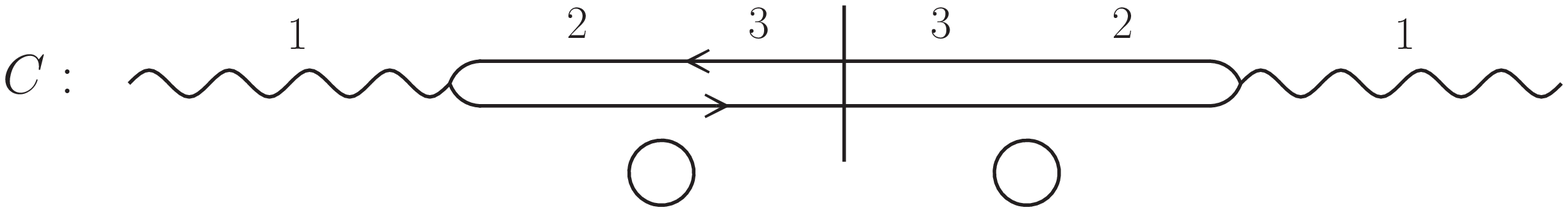}}
  \caption{Gluon $\rightarrow$ quark-antiquark  production in proton-nucleus collisions. The circles represent the nuclei. The numbers above the Feynman diagrams label the various places that a gluon can be emitted. $(a)$ is a purely initial state interactions; $(c)$ is a purely final state interaction; $(b)$ is a mixture of initial and final state interactions.}
  \label{fig:twojets}
\end{figure}
\begin{figure}[h]
  \centering
  \subfigure[]{\label{fig:cto}
  \includegraphics[width=7.5cm]{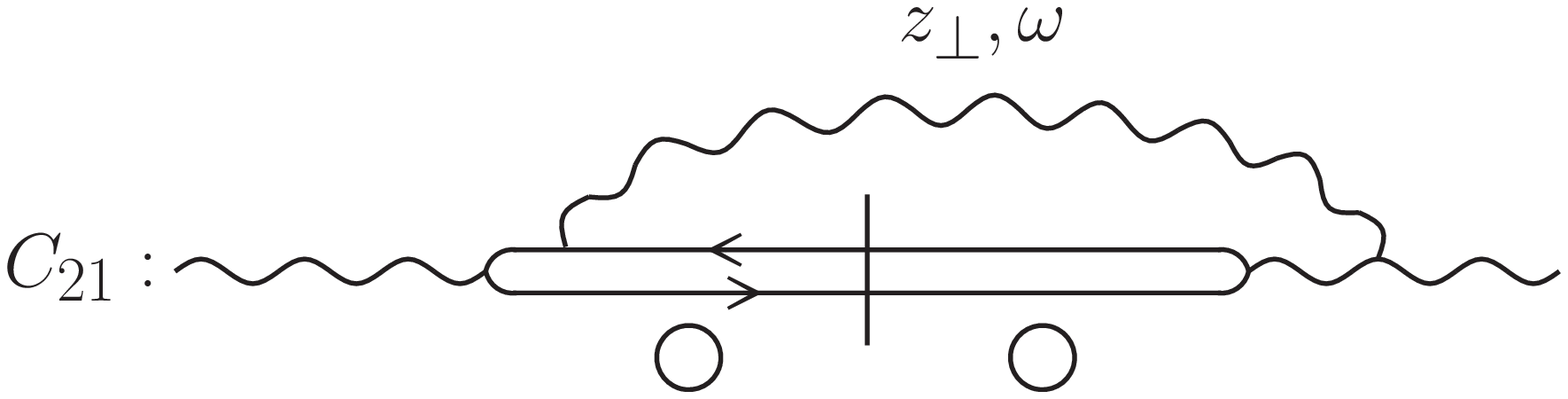}}
  \hspace{0.1in}
  \subfigure[]{\label{fig:aoo}
  \includegraphics[width=7.5cm]{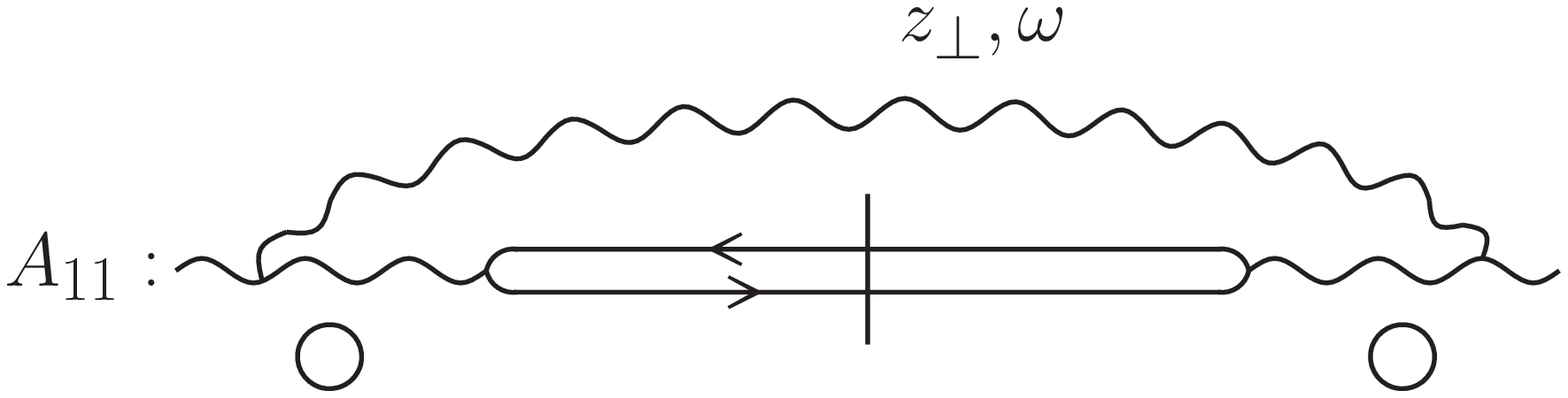}}
  \caption{Two examples for the graphs in \cref{fig:twojets}. $(a)$ and $(b)$ are typical diagrams from class $C$ and $B$, respectively.}
  \label{fig:expamle}
\end{figure}

Before including gluon radiation there are three classes of graphs in gluon $\rightarrow$ quark-antiquark illustrated in \cref{fig:twojets}. Graphs in class $A$, \cref{fig:a}, correspond to interaction of the gluon initiating the process with the nucleus indicated by the small circle in the graph. Graphs in class $C$, \cref{fig:c}, have the quark-antiquark interacting with the nucleus while in class $B$, \cref{fig:b}, the quark-antiquark in the amplitude and the gluon in the complex conjugate amplitude interact with the nucleus. Graphs in class $B$ will be multiplied by a factor of 2 to account for gluon interactions in the amplitude and quark-antiquark interactions in the complex conjugate amplitude. Finally, the numbers just above the graphs indicate the emission points of the radiative gluon which are used to characterize the emission. There are three possible emission points in the amplitude, to the left of the vertical cut, and three possible emission points in the complex conjugate amplitude, to the right of the cut. Emission points to the left (right) of the nucleus are initial state (final state) emissions in the amplitude while in the complex conjugate amplitude initial state (final state) emissions occur to the right (left) of the nucleus. Each gluon emission graph is characterized by a letter, indicating the class, and by two numbers, the first number indicating the emission point in the amplitude and the second number the emission point in the complex conjugate amplitude. The gluon from the proton carries an energy $E$ with transverse coordinate zero both in the amplitude and complex conjugate amplitude. The energy of the radiated gluon is $\omega$ and its transverse coordinate is $z_{\perp}$. For example, $C_{21}$ is shown in \cref{fig:cto}.

Now let us begin to evaluate the graphs. Starting with single purely initial state emission in class $A$
\begin{equation}
  A_{11}=\frac{\alpha_{s}N_{c}}{\pi^{2}}\int\frac{d^{2}z_{\perp}}{z_{\perp}^{2}}
\end{equation}
where the graph, in detail, is shown in \cref{fig:aoo}. The integration limits on $z_{\perp}$ will be given later as will the details of our normalization. We note that interactions of the gluons passing over the nucleus cancel between inelastic and elastic scatterings off nucleons in the nucleus. To achieve the cancellation we allow the case where there may be no interaction, either of the gluon $(\omega,z_{\perp})$ or of the gluon $(E,x_{\perp}=0)$, with nucleons in the amplitude, in the complex conjugate amplitude, or in both the amplitude and complex conjugate amplitude. Such ``noninteraction'' terms ultimately are canceled after all graphs are taken into account. In class $B$ there are two purely initial state radiation contributions,
\begin{subequations}
  \begin{align}
    \label{eq:boo}
    B_{11}&=-\frac{\alpha_{s}N_{c}}{\pi^{2}}\int \frac{d^{2}z}{z^{2}}\, S(x/2)S(-x/2),\\
    \label{eq:bto}
    B_{21}&=-\frac{\alpha_{s}N_{c}}{2\pi^{2}}\int d^{2}z\bigg[\frac{z\cdot (z-x/2)}{z^{2}(z-x/2)^{2}}+\frac{z\cdot (z+x/2)}{z^{2}(z+x/2)^{2}}\bigg]S(x/2)S(-x/2)
  \end{align}
  \label{eq:bb}
\end{subequations}
where
\begin{equation}
  S(x_{\perp})=e^{-Q_{s}^{2}x_{\perp}^{2}/8}
\end{equation}
is the $S$ matrix for the scattering of a quark dipole on the nucleus in the large $N_{c}$ approximation and $Q_{s}$ is the gluon saturation momentum
\begin{equation}
  Q_{s}^{2}=\frac{4\pi^{2}\alpha_{s}N_{c}}{N_{c}^{2}-1}\rho Lx G(x,1/x_{\perp}^{2}).
\end{equation}
For notational simplicity we have dropped the $\perp$ symbol on transverse quantities in \cref{eq:bb}, which we will continue to do in the following if it does not lead to confusion.

Finally, for graphs in class $C$
\begin{subequations}
  \begin{align}
    \label{eq:coo}
    C_{11}&=\frac{\alpha_{s}N_{c}}{\pi^{2}}\int \frac{d^{2}z}{z^{2}},\\
    \label{eq:ctt}
    C_{22}&=\frac{\alpha_{s}N_{c}}{2\pi^{2}}\int d^{2}z\bigg[\frac{1}{(z-x/2)^{2}}+\frac{1}{(z+x/2)^{2}}\bigg],
  \end{align}
\end{subequations}
while $C_{12}$ and $C_{21}$ do not have logarithmic domains of integration in $z_{\perp}$ and are neglected. The integrands in \cref{eq:bto,eq:ctt} can be simplified to
\begin{equation}
  \frac{1}{(z-x/2)^{2}}+\frac{1}{(z+x/2)^{2}}\simeq \frac{2}{z^{2}}\bigg(1+\frac{x^{2}}{4z^{2}}\bigg)
  \label{eq:expandone}
\end{equation}
and
\begin{equation}
  \frac{z}{z^{2}}\cdot \bigg[\frac{z-x/2}{(z-x/2)^{2}}+\frac{z+x/2}{(z+x/2)^{2}}\bigg]\simeq\frac{2}{z^{2}}
  \label{eq:expandtwo}
\end{equation}
where we keep all terms of zeroth order or of first order in $x^{2}$. (We recall that $x^{2}\equiv x_{\perp}^{2}\sim 1/M^{2}$ with $M$ the transverse momentum of each jet, and we always assume $Q_{s}^{2}/M^{2}\ll 1$.)

The domains where logarithmic integration occurs are easily determined and given below with the various contributions having logarithmic integration listed in the corresponding domain:
\begin{subequations}
  \begin{align}
    \frac{1}{M^{2}}<z_{\perp}^{2}<\frac{E}{\omega M^{2}}\ &: \ A_{11},\, B_{21},\, C_{22},\\
    \frac{E}{\omega M^{2}}<z^{2}_{\perp}<\bigg(\frac{E}{\omega M}\bigg)^{2}\ &:\ A_{11},\, B_{11},\, C_{11},\\
    \bigg(\frac{E}{\omega M}\bigg)^{2}<z_{\perp}^{2}<\frac{1}{\mu^{2}}\ &:\ A_{11},\, B_{11},\, C_{11}.
  \end{align}
  \label{eq:initialps}
\end{subequations}
One finds
\begin{equation}
  A_{11}+2B_{21}+C_{22}=\frac{2\alpha_{s}N_{c}}{\pi}\int^{E/\omega M^{2}}_{1/M^{2}}\frac{d z^{2}}{z^{2}}\bigg[1-e^{-Q_{s}^{2} x^{2}/16}+\frac{x^{2}}{8z^{2}}\bigg],
  \label{eq:initialabc}
\end{equation}
and 
\begin{equation}
  A_{11}+2B_{11}+C_{11}=\frac{2\alpha_{s}N_{c}}{\pi}\int^{1/\mu^{2}}_{E/\omega M^{2}}\frac{d z^{2}}{z^{2}}\big(1-e^{-Q_{s}^{2}x^{2}/16}\big).
  \label{eq:initialabcone}
\end{equation}
In the end we shall expand $e^{-Q_{s}^{2}x^{2}/16}$ up to first order in $Q_{s}^{2}x^{2}$ but for the moment we keep the full exponential form. We note that the contributions, $A_{11}$, $B_{11}$ and $C_{11}$, in the domains where $z_{\perp}^{2}>E/\omega M^{2}$ should not be included here since that region of gluon emission is already in the gluon distribution of the incident proton, however, since all initial state interactions will cancel when we form the difference between the proton-nucleus and the proton-proton two-jet energy spectrum we can keep these terms without danger of double counting. 

\subsection{Purely final state radiation}
We begin our calculation of purely final state radiation following what has been done for initial state radiation. Thus in the region $1/M^{2}<z_{\perp}^{2}<E/\omega M^{2}$ corresponding to times after the collision $\omega z_{\perp}^{2} \lesssim E/M^{2}$ we have final state contributions, analogous to those given in \cref{eq:initialps},
\begin{equation}
  \frac{1}{M^{2}}<z_{\perp}^{2}<\frac{E}{\omega M^{2}}\ :\ A_{22},\, B_{32},\, C_{33}.
\end{equation}
These contributions are
\begin{subequations}
  \begin{align}
    A_{22}&=\frac{\alpha_{s}N_{c}}{\pi^{2}}\int \frac{d^{2}z}{z^{2}},\\
    B_{32}&=-\frac{\alpha_{s}N_{c}}{2\pi^{2}}\int d^{2}z\, \frac{z}{z^{2}}\cdot \bigg[\frac{z-x/2}{(z-x/2)^{2}}+\frac{z+x/2}{(z+x/2)^{2}}\bigg]S(x/2)S(-x/2),\\
    C_{33}&=\frac{\alpha_{s}N_{c}}{2\pi^{2}}\int d^{2}z\, \bigg[\frac{1}{(z-x/2)^{2}}+\frac{1}{(z+x/2)^{2}}\bigg].
  \end{align}
\end{subequations}
Using \cref{eq:expandone,eq:expandtwo} one finds
\begin{equation}
  A_{22}+2B_{32}+C_{33}=\frac{2\alpha_{s}N_{c}}{\pi}\int^{E/\omega M^{2}}_{1/M^{2}}\frac{d z^{2}}{z^{2}}\bigg(1-e^{-Q_{s}^{2}x^{2}/16}+\frac{x^{2}}{8z^{2}}\bigg).
  \label{eq:finalabc}
\end{equation}
When $z^{2}>E/\omega M^{2}$, or equivalently $\omega z^{2}>E /M^{2}$, one can no longer view the quark-antiquark jet pair as having a frozen transverse coordinate. After the scattering the separation between the quark and antiquark, $\Delta x_{\perp}$ is given by
\begin{equation}
  \Delta x_{\perp} \simeq x_{\perp}+\omega z^{2}\cdot \frac{M}{E}=x_{\perp}+\frac{\omega z^{2}}{E/M^{2}}\cdot \frac{1}{M}.
\end{equation}
Since $x_{\perp}$ is of size $1/M$ we see that when $\omega z^{2}\gg E/M^{2}$ $\Delta x_{\perp}$ will generally be much larger than $1/M$. This is the region where the decays of the jets become important. We define the quark and antiquark jets by cones of half-angle $\delta$ centered about their directions of motion. The gluon radiation spectrum additional to \cref{eq:finalabc} is then given by radiation coming off the quark (or antiquark) at an angle $\theta$ in the interval
\begin{equation}
  \delta^{2}<\theta^{2}<\frac{M^{2}}{E \omega}
\end{equation}
where the upper limit of $\theta^{2}$ is determined by requiring the gluon formation time be equal to $E/M^{2}$, that is $1/\omega \theta_{\textrm{max}}^{2}=E/M^{2}$. Gluons emitted at large angle are already included in \cref{eq:finalabc}. Thus
\begin{equation}
  A_{33}+C_{33}=\frac{2\alpha_{s}N_{c}}{\pi}\int^{M^{2}/E\omega}_{\delta^{2}}\frac{d \theta^{2}}{\theta^{2}}
\end{equation}
while
\begin{equation}
  2B_{33}=-\frac{2\alpha_{s}N_{c}}{\pi}\int^{M^{2}/E \omega}_{\delta^{2}}\frac{d \theta^{2}}{\theta^{2}}\, S^{2}(x/2).
\end{equation}
It is convenient to write this as
\begin{equation}
  A_{33}+2B_{33}+C_{33}=\frac{2\alpha_{s}N_{c}}{\pi}\int^{(M/E)^{2}}_{\delta^{2}}\frac{d \theta^{2}}{\theta^{2}}\big(1-e^{-Q_{s}^{2}x^{2}/16}\big)+\frac{2\alpha_{s}N_{c}}{\pi}\int^{(E/\omega M)^{2}}_{E/\omega M^{2}}\frac{d z^{2}}{z^{2}}\big(1-e^{-Q_{s}^{2}x^{2}/16}\big).
  \label{eq:finalcom}
\end{equation}
If the produced quark-antiquark pair are heavy quarks having mass $M$ then the first term on the right-hand side of \cref{eq:finalcom} is absent, the $\theta$ integration going over the ``dead cone'' \cite{Dokshitzer}. \Cref{eq:finalabc,eq:finalcom} are our final results for final state energy radiation. We note that \cref{eq:initialabc,eq:finalabc} are identical. Also \cref{eq:initialabcone} and the second term on the right-hand side of \cref{eq:finalcom} are identical except for upper limits on the $z_{\perp}$ integration reflecting the fact that the final state radiation is off jets.

\subsection{Initial-state-final-state interference terms}
The interference terms are straightforward to calculate, but there are many terms. We shall first give expression for the various terms and then their domains. Thus,
\begin{subequations}
  \begin{align}
    A_{12}=A_{21}&=-\frac{\alpha N_{c}}{\pi^{2}}\int \frac{d^{2}z}{z^{2}}\, S^{2}(z),\\
    \label{eq:ifa}
    A_{13}=A_{31}&=-\frac{\alpha_{s}N_{c}}{2\pi^{2}}\int d^{2}z\,\frac{z}{z^{2}}\cdot \bigg[\frac{z-x/2}{(z-x/2)^{2}}+\frac{z+x/2}{(z+x/2)^{2}}\bigg]S^{2}(z).
  \end{align}
\end{subequations}
For type $C$ graphs
\begin{subequations}
  \begin{align}
    C_{23}=C_{32}&=-\frac{\alpha_{s}N_{c}}{2\pi^{2}}\int d^{2}z\, \bigg[\frac{S^{2}(z-x/2)}{(z-x/2)^{2}}+\frac{S^{2}(z+x/2)}{(z+x/2)^{2}}\bigg],\\
    \label{eq:ifc}
    C_{13}=C_{31}&=-\frac{\alpha_{s}N_{c}}{2\pi^{2}}\int d^{2}z\, \frac{z}{z^{2}}\cdot \bigg[\frac{z+x/2}{(z+x/2)^{2}}S^{2}(z+x/2)+\frac{z-x/2}{(z-x/2)^{2}}S^{2}(z-x/2)\bigg].
  \end{align}
\end{subequations}
Finally, for class $B$ graphs
\begin{align}
  B_{22}=B_{13}=B_{31}&=\frac{\alpha N_{c}}{2\pi^{2}}\int d^{2}z\,\frac{z}{z^{2}}\cdot \bigg[\frac{z+x/2}{(z+x/2)^{2}}S(z)S(x/2)S(z+x/2)\nonumber\\
  &\phantom{====}+\frac{z-x/2}{(z-x/2)^{2}}S(z)S(x/2)S(z-x/2)\bigg],
  \label{eq:ifb}
\end{align}
are the only nonzero contributions. The different region for these contributions are
\begin{subequations}
  \begin{align}
    \textrm{region}& \ 2,\quad \frac{1}{M^{2}}<z^{2}<\frac{E}{\omega M^{2}}:\quad A_{12},\, A_{21},\, C_{23},\, C_{32},\, B_{22},\, B_{31};\\
    \textrm{region}& \ 1,\quad \frac{E}{\omega M^{2}}<z^{2}<\bigg(\frac{E}{\omega M}\bigg)^{2}:\quad A_{13},\, A_{31},\, C_{13},\, C_{31},\, B_{13},\, B_{31}.
  \end{align}
  \label{eq:ifregion}
\end{subequations}
In order to further evaluate these contributions it is useful to expand the $S$ matrices appearing in \cref{eq:ifa,eq:ifc,eq:ifb} in the small-$x$ limit. Thus
\begin{subequations}
  \begin{align}
  S^{2}(z-x/2)&\simeq e^{-Q_{s}^{2}z^{2}/4}\bigg(1+\frac{Q_{s}^{2}}{4}z\cdot x-\frac{Q_{s}^{2}}{16}x^{2}+\cdots \bigg),\\
  S(z)S(x/2)S(z-x/2)&\simeq e^{-Q_{s}^{2}z^{2}/4}\bigg(1+\frac{Q_{s}^{2}}{8}z\cdot x-\frac{Q_{s}^{2}}{16}x^{2}+\cdots \bigg).
\end{align}
\label{eq:expandthree}
\end{subequations}
In the region $E/\omega M^{2}<z^{2}<(E/\omega M)^{2}$ the total initial-final interference contribution is
\begin{equation}
  (A+2B+C)_{1}=-\frac{2\alpha_{s}N_{c}}{\pi^{2}}\int d^{2}z\, \frac{z}{z^{2}}\cdot \frac{(z-x/2)}{(z-x/2)^{2}}\Big[ S^{2}(z)+S^{2}(z-x/2)-2S(z)S(x/2)S(z-x/2)\Big]
  \label{eq:crossfullone}
\end{equation}
where the subscript indicates the region $E/\omega M^{2}<z^{2}<(E/\omega M)^{2}$. Using \cref{eq:expandthree} one gets 
\begin{equation}
  S^{2}(z)+S^{2}(z-x/2)-2S(z)S(x/2)S(z-x/2)\simeq S^{2}(z)\frac{Q_{s}^{2}}{16}x^{2}
  \label{eq:qqexpand}
\end{equation}
giving 
\begin{equation}
  (A+2B+C)_{1}=-\frac{\alpha_{s}N_{c}Q_{s}^{2}x^{2}}{8\pi}\int^{(E/M\omega)^{2}}_{E/\omega M^{2}}\frac{dz^{2}}{z^{2}}\, e^{-Q_{s}^{2}z^{2}/4}.
  \label{eq:crossone}
\end{equation}
In the region $1/M^{2}<z^{2}<E/\omega M^{2}$, indicated by the subscript 2 below, the contribution is
\begin{equation}
  (A+2B+C)_{2}=-\frac{2\alpha_{s} N_{c}}{\pi^{2}}\int d^{2}z\bigg[\frac{S^{2}(z)}{z^{2}}+\frac{S^{2}(z-x/2)}{(z-x/2)^{2}}-2\frac{z\cdot(z-x/2)}{z^{2}(z-x/2)^{2}}S(x/2)S(z)S(z-x/2)\bigg].
  \label{eq:crossfulltwo}
\end{equation}
Using \cref{eq:expandthree} one finds
\begin{equation}
  (A+2B+C)_{2}=-\frac{\alpha_{s}N_{c}Q_{s}^{2}x^{2}}{4\pi}\int^{E/\omega M^{2}}_{1/M^{2}}\frac{dz^{2}}{z^{2}}e^{-Q_{s}^{2}z^{2}/4}\bigg(1+\frac{2}{Q_{s}^{2}z^{2}}\bigg).
  \label{eq:crosstwo}
\end{equation}

\subsection{Putting all the terms together}
\Cref{eq:initialabc,eq:initialabcone,eq:finalabc,eq:finalcom,eq:crossone,eq:crosstwo} give the final results for gluon radiation occurring during the hard process gluon $+$ nucleus $\rightarrow$ quark-antiquark jets. Our real interest is in the change in the per event gluon radiation in going from a proton target to a nuclear target. Without gluon radiation the three graphs $A,B,C$ of \cref{fig:twojets} give
\begin{equation}
  (A+2B+C)_{\textrm{no radiation}}=1+1-2e^{-Q_{s}^{2}x^{2}/16}\simeq \frac{Q_{s}^{2}x^{2}}{8}.
  \label{eq:norad}
\end{equation}
If we divide our above results by \cref{eq:norad} we get a per event radiation spectrum.

We note some unusual terms, proportional to $d^{2}z/z^{4}$, which do not vanish when $Q_{s}^{2}$ is set to zero. These terms are present in \cref{eq:initialabc,eq:finalabc,eq:crosstwo}. A term which survives as $Q_{s}^{2}\rightarrow 0$ is a term where no scattering with the target occurs. Such terms should not be present in our final answer and, indeed, they cancel when the contributions of \cref{eq:initialabc,eq:finalabc,eq:crosstwo} are added together. After this cancellation is accounted for, all remaining terms go to zero as $Q_{s}^{2}x^{2}$ when $x^{2}\rightarrow 0$. We keep only the linear term in $Q_{s}^{2}x^{2}$, appropriate to the case where $Q_{s}^{2}/M^{2}\ll 1$, but we keep all powers in $z^{2}Q_{s}^{2}$. In order to evaluate the difference of the proton-nucleus and proton-proton spectra we first divide all results by $Q_{s}^{2}x^{2}/8$, as given in \cref{eq:norad}, to get a normalized spectrum. The resulting initial and final state scattering terms then have no $Q_{s}^{2}$ dependence whatsoever so they cancel in the difference between nuclear and proton targets. The interference terms given in \cref{eq:crossone,eq:crosstwo} give a nonzero contribution to the difference between the spectrum for proton and nuclear targets in the region
\begin{equation}
  \frac{4}{Q_{s}^{2}}<z^{2}<\textrm{min}\bigg[\bigg(\frac{E}{M\omega}\bigg)^{2},\frac{4}{Q_{s}^{2}(\textrm{P})}\bigg],
  \label{eq:apregion}
\end{equation}
where the factors of 4 in \cref{eq:apregion} represent our best guess as to how numerical limits to the logarithmic integrals come in (see \cref{app:ps}). $Q_{s}^{2}$ is, as always, the saturation momentum of the nucleus, and $Q_{s}^{2}(\textrm{P})$ is the saturation momentum of the proton if $x$ values are small enough for the proton saturation momentum to have meaning, otherwise $Q_{s}(\textrm{P})$ should be taken to be a hadron scale of, say, 300 to 500$\,\textrm{MeV}$. Dividing \cref{eq:crossone,eq:crosstwo} by \cref{eq:norad} and using the resulting expression in the region \cref{eq:apregion} gives
\begin{equation}
  \omega\frac{d I^{\textrm{Nucleus}-\textrm{Proton}}}{d\omega}=\frac{\alpha_{s}N_{c}}{\pi}\ln \frac{E^{2}Q_{s}^{2}}{4M^{2}\omega^{2}}=\frac{2\alpha_{s}N_{c}}{\pi}\ln\frac{EQ_{s}}{2M\omega},
  \label{eq:radspec}
\end{equation}
in case $M\omega /E>Q_{s}(\textrm{P})/2$. In case $M \omega/E <Q_{s}(\textrm{P})/2$ one has
\begin{equation}
  \label{eq:radspecqs}
  \omega\frac{dI}{d\omega}=\frac{\alpha_{s}N_{c}}{\pi}\ln\frac{Q_{s}^{2}}{Q_{s}^{2}(\textrm{P})},
\end{equation}
a result which does not depend on $\omega$. \Cref{eq:radspec,eq:radspecqs} are exactly the result of Arleo and Peign\'e \cite{Arleo}.

\subsection{Small-$x$ evolution}
\begin{figure}[h]
  \centering
  \subfigure[]{\label{fig:dipoleone}
  \includegraphics[width=7.5cm]{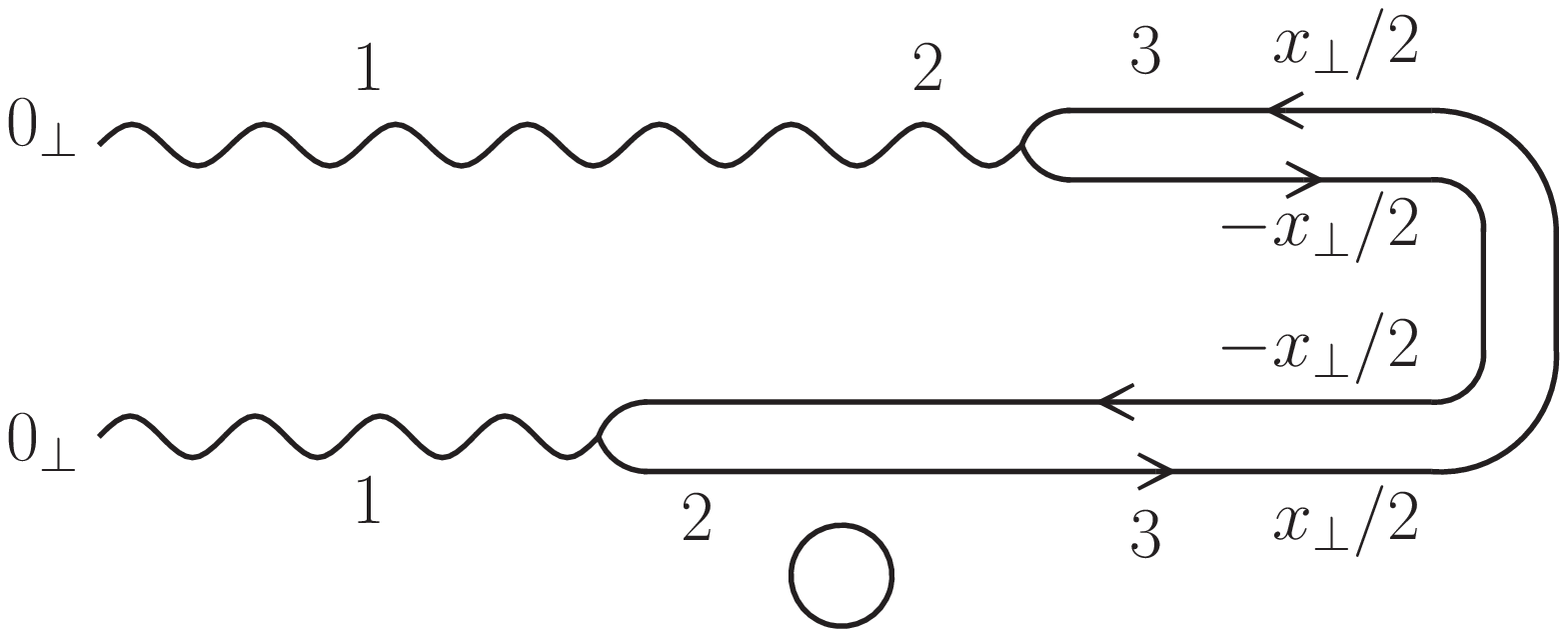}}
  \hspace{0.1in}
  \subfigure[]{\label{fig:dipolerad}
  \includegraphics[width=7.5cm]{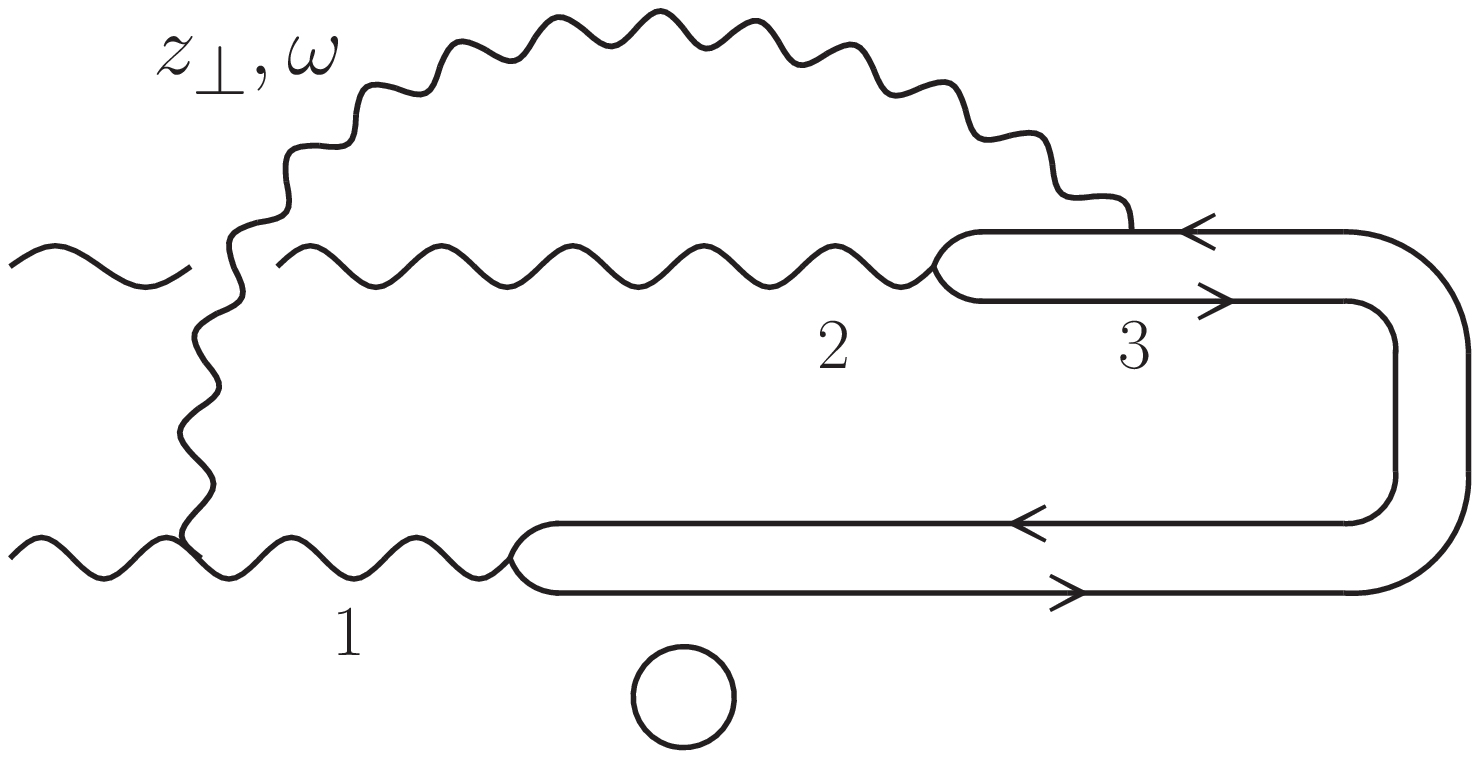}}
  \caption{Two-jet production with small-$x$ evolution. $(a)$ is a cross diagram without a gluon radiation in the dipole form. $(b)$ is the energy loss diagram.}
  \label{fig:evolution}
\end{figure}

So far we have treated the scattering of the incoming gluon, two gluon, quark-antiquark, or quark-antiquark gluon on the nucleus in a simple Glauber approximation. However, it is easy to include QCD small-$x$ evolution in the Gaussian approximation \cite{Iancu} so long as $Q_{s}/M\ll 1$. (Small-$x$ evolution in general is not difficult to implement although numerical evaluations of dipole scattering amplitudes are required for explicit results. See below.) The result in the Gaussian approximation is that one uses the formulas exactly as we have given them but $Q_{s}^{2}$ and $Q_{s}^{2}(\textrm{P})$ are taken to be $x$ dependent with $x_{\textrm{BJ}}\simeq M^{2}/2Em_{\textrm{p}}$ [see \cref{eq:smallxregion} for a more precise range of $x$ values] where $m_{\textrm{p}}$ is the proton mass and $M$ and $E$ are the jet transverse momentum (heavy quark mass) and $E$ the energy of the jets or heavy quarks. Let us briefly indicate how this comes about. It should suffice to take graphs in class $B$ as shown in \cref{fig:twojets}. If we view this as an amplitude rather then as a cut amplitude, which should be valid for leading and next-to-leading small-$x$ evolution \cite{Mueller}, the amplitude can be illustrated as in \cref{fig:dipoleone}. The complex conjugate amplitude in \cref{fig:twojets} appears as the upper half of the graph in \cref{fig:dipoleone}. The first, highest energy $\omega$, gluon emission which accounts for the energy loss must connect the lower part of the graph in \cref{fig:dipoleone} to the upper part of that graph. Subsequent gluon evolution involving gluons having energy less than $\omega$ can appear anywhere in the graph. If the first gluon $\omega$ is a purely initial state emission going from regions 1 or 2 in the lower part of the graph to region 1 in the upper part of the graph then this first emission has no effect on the subsequent softer emissions. Exactly the same is the case for purely final state interactions. Thus, when the first gluon emission is purely initial state or purely final state the evolution is exactly as if there were no emission and thus these terms will continue to cancel when nucleus minus proton differences are taken, and this cancellation does not depend on the Gaussian approximation.

Now consider initial-state-final-state interference graphs. As an example we take the $B_{13}$ term evaluated earlier and illustrated in the dipole formalism in \cref{fig:dipolerad}. The system passing over the nucleus in \cref{fig:dipolerad} consists of a gluon $(z_{\perp},\omega)$, a gluon $(0_{\perp},E)$, a quark $(x_{\perp}/2,E/2)$ and an antiquark $(-x_{\perp}/2,E/2)$. Because of the large $N_{c}$ approximation this two gluon-quark-antiquark systems can be viewed as three independent (quark,antiquark) dipoles at positions $(-x_{\perp}/2,0_{\perp})$, $(0_{\perp},z_{\perp})$, $(z_{\perp},x_{\perp}/2)$. The scattering has the factor given by the last factor on the right-hand side of \cref{eq:crossfullone}. If we take the $S$ matrices to be given by Balitsky-Kovchegov evolution (BK) \cite{Balitsky,Kovchegov} starting from a McLerran-Venugopalan initial condition then \cref{eq:crossfullone,eq:crossfulltwo} are general, including evolution, in the large $N_{c}$ approximation. However, \cref{eq:expandthree,eq:qqexpand} require the Gaussian approximation.

Thus in the Gaussian approximation we expect \cref{eq:crossone,eq:crosstwo} and hence \cref{eq:radspec,eq:radspecqs} to have small-$x$ evolution included in $Q_{s}$ and $Q_{s}(\textrm{P})$ while more generally small-$x$ evolution is included, via BK evolution, in \cref{eq:crossfullone,eq:crossfulltwo} when the $S$ matrices are evaluated using BK evolution. In the Gaussian approximation the evolution results in a saturation momentum depending on $1/x_{\textrm{BJ}}=2\omega m_{\textrm{p}}z_{\perp}^{2}$. For a fixed $\omega$ and $1/Q^{2}_{s}<z_{\perp}^{2}<(E/M \omega)^{2}$ (see \cref{eq:apregion,eq:radspec}) this leads to $x_{\textrm{BJ}}$ values in the region
\begin{equation}
  \label{eq:smallxregion}
  \frac{2\omega m_{\textrm{p}}}{Q_{s}^{2}}<\frac{1}{x_{\textrm{BJ}}}<\frac{E}{\omega}\frac{2Em_{\textrm{p}}}{M^{2}}.
\end{equation}

\section{Energy loss in quark $\rightarrow$ quark-gluon jets in proton-nucleus collisions\label{sec:qgjet}}
\begin{figure}[h]
  \centering
  \subfigure[]{\label{fig:qga}
  \includegraphics[width=12cm]{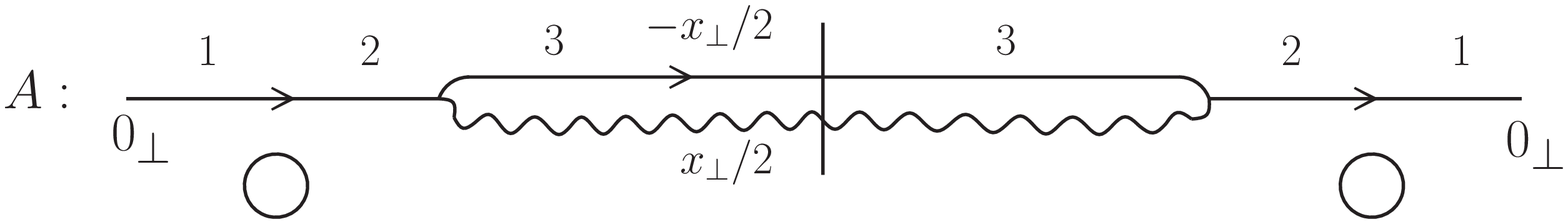}}
  \subfigure[]{\label{fig:qgb}
  \includegraphics[width=12cm]{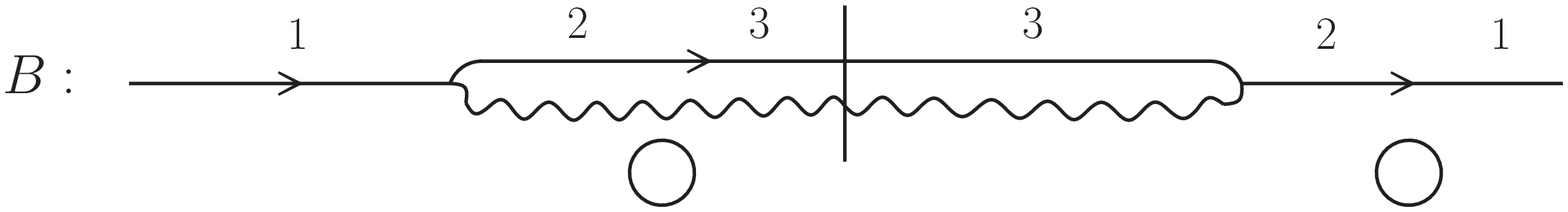}}
  \subfigure[]{\label{fig:qgc}
  \includegraphics[width=12cm]{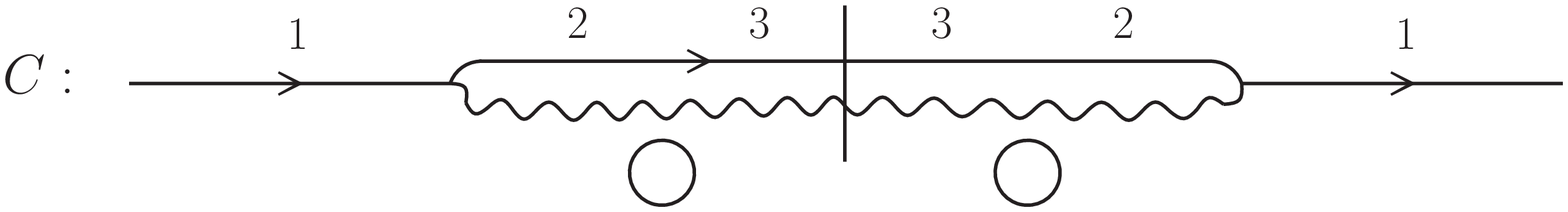}}
  \caption{Quark $\rightarrow$ quark-gluon production in proton-nucleus collisions. The circles represent the nuclei. The numbers above the Feynman diagrams label the various places that a gluon can be emitted. (a), (b) and (c) represent all possible places that the interactions can happen.}
  \label{fig:qgjets}
\end{figure}
This section is pretty much a repeat of what has just been done, but now for quark $\rightarrow$ quark-gluon jets rather than gluon $\rightarrow$ quark-antiquark jets. Of course it would be nice to simply read off the radiative spectrum results for quark $\rightarrow$ quark-gluon in terms of those for gluon $\rightarrow$ quark-antiquark but we, so far, have not been able to find a way to do this. We are thus forced to do another detailed calculation, again in the large $N_{c}$ limit, for the quark initiated process. Analogous to those shown in \cref{fig:twojets} the relevant graphs now are shown in \cref{fig:qgjets}. As in the last section we consider, in turn, purely initial state radiation, purely final state radiation and initial-state-final-state interference terms. We shall be somewhat briefer in our description of the calculations here since the general procedure is exactly as in the previous section.

\subsection{Purely initial state radiation}
\begin{figure}[h]
  \centering
  \includegraphics[width=13cm]{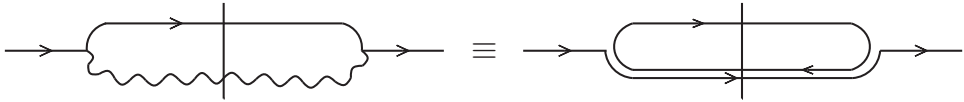}
  \caption{A quark splits into a quark and gluon in the large $N_{c}$ limit.}
  \label{fig:loop}
\end{figure}
The contributions to purely initial state radiation are:
\begin{subequations}
  \begin{align}
    A_{11}&=\frac{\alpha_{s}N_{c}}{2\pi^{2}}\int \frac{d^{2}z}{z^{2}},\\
    \label{eq:qgboo}
    B_{11}&=-\frac{\alpha_{s}N_{c}}{2\pi^{2}}\int \frac{d^{2}z}{z^{2}}S(x)S(x/2),\\
    B_{21}&=-\frac{\alpha_{s}N_{c}}{2\pi^{2}}\int d^{2}z\,\frac{z\cdot (z-x/2)}{z^{2}(z-x/2)^{2}}S(x)S(x/2),\\
    C_{11}&=\frac{\alpha_{s}N_{c}}{2\pi^{2}}\int \frac{d^{2}z}{z^{2}},\\
    \label{eq:qgctt}
    C_{22}&=\frac{\alpha_{s}N_{c}}{2\pi^{2}}\int d^{2}z\, \bigg[\frac{1}{(z-x/2)^{2}}+\frac{x^{2}}{(z-x/2)^{2}(z+x/2)^{2}}\bigg].
  \end{align}
\end{subequations}
The new element here as compared to our previous calculation is the presence of a closed color loop illustrated in \cref{fig:loop} where the gluon line is replaced by a quark-antiquark pair in the large $N_{c}$ limit. Thus for example the $S(x)$ in \cref{eq:qgboo} stands for elastic scattering in the initial state on the closed fermion loop in \cref{fig:loop} while the $S(x/2)$ corresponds to a combination of elastic and inelastic scatterings on the lower fermion line on the right-hand part of \cref{fig:loop}. Similarly the second term on the right-hand side of \cref{eq:qgctt} corresponds to the radiated gluon being emitted and absorbed in the closed fermion (dipole) loop with the result being the usual dipole formula for gluon emission. The kinematic regions where those contributions are important are exactly as in \cref{eq:initialps} of the previous section. Thus one gets
\begin{equation}
  A_{11}+2B_{21}+C_{22}=\frac{\alpha_{s}N_{c}}{\pi}\frac{5Q_{s}^{2}x^{2}}{16}\int^{E/\omega M^{2}}_{1/M^{2}}\frac{dz^{2}}{z^{2}}\bigg[\frac{1}{2}+\frac{2}{Q_{s}^{2}z^{2}}\bigg]
  \label{eq:qgfullabcone}
\end{equation}
for the region $1/M^{2}<z^{2}<E/\omega M^{2}$ while 
\begin{equation}
  A_{11}+2B_{11}+C_{11}=\frac{\alpha_{s}N_{c}}{\pi}\int^{1/\mu^{2}}_{E/\omega M^{2}}\frac{dz^{2}}{z^{2}}\big(1-e^{-\frac{5}{32}Q_{s}^{2}x^{2}}\big)\simeq \frac{\alpha_{s}N_{c}}{\pi}\frac{5Q_{s}^{2}x^{2}}{16}\frac{1}{2}\int^{1/\mu^{2}}_{E/\omega M^{2}}\frac{dz^{2}}{z^{2}}
  \label{eq:qgfullabctwo}
\end{equation}
in the region $E/\omega M^{2}<z^{2}<1/\mu^{2}$ where $\mu$ is an infrared cutoff. As in the previous section the initial state contributions coming from the region $E/\omega M^{2}<z^{2}<1/\mu^{2}$ are already included in the quark parton distribution in the nonradiative two-jet production. In forming the differences between nuclear and proton targets such terms will cancel and the double counting problem will be avoided.

\subsection{Purely final state radiation}
In the region $1/M^{2}<z^{2}<E/\omega M^{2}$ the following contributions are important:
\begin{subequations}
  \begin{align}
    A_{22}&=\frac{\alpha_{s}N_{c}}{2\pi^{2}}\int \frac{d^{2}z}{z^{2}},\\
    B_{32}&=-\frac{\alpha_{s}N_{c}}{2\pi^{2}}\int d^{2}z\, \frac{z\cdot (z-x/2)}{z^{2}(z-x/2)^{2}}S(x)S(x/2),\\
    \label{eq:qgcdtt}
    C_{33}&=\frac{\alpha_{s}N_{c}}{2\pi^{2}}\int d^{2}z\,\bigg[\frac{2}{(z-x/2)^{2}}+\frac{1}{(z+x/2)^{2}}-2\frac{(z-x/2)\cdot(z+x/2)}{(z-x/2)^{2}(z+x/2)^{2}}S^{2}(x)\bigg].
  \end{align}
\end{subequations}
One finds
\begin{equation}
  A_{22}+2B_{32}+C_{33}=\frac{\alpha_{s}N_{c}}{\pi}\frac{5Q_{s}x^{2}}{16}\int^{E/\omega M^{2}}_{1/M^{2}}\frac{dz^{2}}{z^{2}}\bigg[\frac{13}{10}+\frac{2}{Q_{s}^{2}z^{2}}\bigg].
\end{equation}

In the large $z_{\perp}$ region we separate the contribution into two parts. The first part 
\begin{equation}
  \frac{3\alpha_{s}N_{c}}{\pi}\int^{(M/E)^{2}}_{\delta^{2}}\frac{d \theta^{2}}{\theta^{2}}\big[1-S(x)S(x/2)\big]
  \label{eq:qgcone}
\end{equation}
corresponds to emission at very small angles with respect to the jets, but at angle $\theta>\delta$. The prefactor $3\alpha_{s}N_{c}/\pi$ is exactly $\frac{3}{2}$ times that of the first term in \cref{eq:finalcom} reflecting the fact that the color charge squared of the gluon jet is twice that of the quark jet in the large $N_{c}$ approximation. In the region $E/\omega M^{2}<z^{2}<(E/\omega M)^{2}$ one has contributions
\begin{subequations}
  \begin{align}
    A_{33}&=\frac{\alpha_{s}N_{c}}{2\pi^{2}}\int d^{2}z\,\bigg[\frac{1}{(z-x/2)^{2}}+\frac{x^{2}}{(z-x/2)^{2}(z+x/2)^{2}}\bigg],\\
    B_{33}&=-\frac{\alpha_{s}N_{c}}{2\pi^{2}}\int d^{2}z\,\bigg[\frac{1}{(z-x/2)^{2}}+\frac{x^{2}}{(z-x/2)^{2}(z+x/2)^{2}}\bigg]S(x)S(x/2)
  \end{align}
\end{subequations}
and $C_{33}$ as given in \cref{eq:qgcdtt}. Including the contribution \cref{eq:qgcone} one gets a total contribution for $z^{2}>E/\omega M^{2}$
\begin{equation}
  A_{33}+2B_{33}+C_{33}=\frac{3\alpha_{s}N_{c}}{\pi}\int^{(M/E)^{2}}_{\delta^{2}}\frac{d \theta^{2}}{\theta^{2}}\big(1-e^{-\frac{5}{32}Q_{s}^{2}x^{2}}\big)+\frac{\alpha_{s}N_{c}}{\pi}\frac{5Q_{s}^{2}x^{2}}{16}\frac{13}{10}\int^{(E/\omega M)^{2}}_{E/\omega M^{2}}\frac{d z^{2}}{z^{2}}.
  \label{eq:qgtotalabc}
\end{equation}

\subsection{Initial-state-final-state interference terms}
The interference terms are
\begin{subequations}
  \begin{align}
    A_{12}=A_{21}&=-\frac{\alpha_{s}N_{c}}{2\pi^{2}}\int \frac{d^{2}z}{z^{2}}S^{2}(z),\\
    \label{eq:acbbb}
    \begin{pmatrix}
    A_{13}=A_{31}\\
    C_{13}=C_{31}\\
    B_{22}=B_{13}=B_{31}
    \end{pmatrix}
  &=-\frac{\alpha_{s}N_{c}}{2\pi^{2}}\int d^{2}z\,\frac{z\cdot (z-x/2)}{z^{2}(z-x/2)^{2}}
  \begin{pmatrix}
    S^{2}(z)\\
    S^{2}(z-x/2)\\
    -S(x)S(z)S(z-x/2)
  \end{pmatrix},\\
  C_{23}=C_{32}&=-\frac{\alpha_{s}N_{c}}{2\pi^{2}}\int d^{2}z\,\bigg[\frac{2S^{2}(z-x/2)}{(z-x/2)^{2}}+\frac{S^{2}(z+x/2)}{(z+x/2)^{2}}\\
    &\phantom{==}-\frac{(z+x/2)\cdot (z-x/2)}{(z+x/2)^{2}(z-x/2)^{2}}\big(S^{2}(z+x/2)+S^{2}(z-x/2)\big)\bigg].
  \end{align}
  \label{eq:qgacb}
\end{subequations}
Keeping only linear term in $x^{2}$ one finds
\begin{equation}
  (A+2B+C)_{2}=-\frac{\alpha_{s}N_{c}}{\pi}\frac{5Q_{s}^{2}x^{2}}{16}\int^{E/\omega M^{2}}_{1/M^{2}}\frac{d z^{2}}{z^{2}}\, e^{-Q_{s}^{2}z^{2}/4}\bigg[\frac{9}{5}+\frac{4}{Q_{s}^{2}z^{2}}\bigg]
  \label{eq:qgregiontwo}
\end{equation}
and
\begin{equation}
  (A+2B+C)_{1}=-\frac{\alpha_{s}N_{c}}{\pi}\frac{5Q_{s}^{2}x^{2}}{16}\frac{4}{5}\int^{(E/\omega M)^{2}}_{E/\omega M^{2}}\frac{dz^{2}}{z^{2}}e^{-Q_{s}^{2}z^{2}/4}.
  \label{eq:qgregionone}
\end{equation}
The subscripts 1 and 2 on $(A+2B+C)$ indicate the same combination of terms as in \cref{eq:ifregion} of the previous section. We have isolated the factor $5Q_{s}^{2}x^{2}/16$ in front of \cref{eq:qgregiontwo,eq:qgregionone} because that is the factor that appears when there is no radiation, analogous to \cref{eq:norad} in the previous section, and so dividing our formulas by this factor will give the radiation spectrum $\omega dI/d\omega$.

In evaluating the change in the spectrum in going from a proton to a nuclear target initial and final state emissions cancel leaving only the initial-state-final-state interference terms given in \cref{eq:qgregiontwo,eq:qgregionone}. Keeping linear terms in $Q_{s}^{2}x^{2}$ and dividing by $5Q_{s}^{2}x^{2}/16$ we find, in analogy with \cref{eq:radspec,eq:radspecqs}
\begin{equation}
  \omega\frac{d I^{\textrm{Nucleus}-\textrm{Proton}}}{d\omega}=\frac{8}{5}\frac{\alpha_{s}N_{c}}{\pi}\ln\frac{EQ_{s}}{2M\omega}
  \label{eq:radspecqgone}
\end{equation}
for $(\omega M/E)>Q_{s}(\textrm{P})/2$ and 
\begin{equation}
  \label{eq:radspecqgtwo}
  \omega\frac{dI^{\textrm{Nucleus}-\textrm{Proton}}}{d\omega}=\frac{4}{5}\frac{\alpha_{s}N_{c}}{\pi}\ln\frac{Q^{2}_{s}}{Q^{2}_{s}(P)}
\end{equation}
for $(\omega M/E)<Q_{s}(\textrm{P})/2$. We note that \cref{eq:radspecqgone,eq:radspecqgtwo} are surprisingly close to \cref{eq:radspec,eq:radspecqs} a fact (coincidence ?) for which we have no explanation.

\subsection{Rough estimates of energy loss}
Now let us use \cref{eq:radspecqgone,eq:radspecqgtwo} to get some rough estimates of the typical energy loss difference in an event on a nuclear target and on a proton target. A natural definition of typical energy loss $\bar{\omega}$ is to require
\begin{equation}
  \int_{\bar{\omega}}d\omega\, \frac{d I^{\textrm{Nucleus}-\textrm{Proton}}}{d\omega}=\frac{1}{2}.
  \label{eq:avg}
\end{equation}
We first use \cref{eq:avg} to give an estimate of $\bar{\omega}$ at RHIC energies for forward, quark initiated, two-jet production. If $(\bar{\omega}M/E)>Q_{s}(\textrm{P})/2$ one can use \cref{eq:radspecqgone} in \cref{eq:avg} to get 
\begin{equation}
  \frac{8\alpha_{s}N_{c}}{5\pi}\ln^{2}\frac{EQ_{s}}{2M \bar{\omega}}=1
\end{equation}
or
\begin{equation}
  \frac{E}{\bar{\omega}}=\frac{2M}{Q_{s}}\exp\bigg(\sqrt{\frac{5\pi}{8\alpha_{s} N_{c}}}\bigg).
  \label{eq:eo}
\end{equation}
Taking $M=3\,\textrm{GeV}$, $Q_{s}=1.5\,\textrm{GeV}$, $\alpha_{s}=\frac{2}{5}$ one finds
\begin{equation}
  \bigg(\frac{E}{\bar{\omega}}\bigg)_{\textrm{RHIC}}\simeq 14.
  \label{eq:rhic}
\end{equation}
$(2M\bar{\omega})/E\simeq \frac{3}{7}$ so that for $Q_{s}(\textrm{P})\lesssim \frac{1}{2}\, \textrm{GeV}$ the condition for using \cref{eq:radspecqgone} is satisfied. For LHC energies take $M=7\,\textrm{GeV}$, $Q_{s}=2.5\, \textrm{GeV}$ and $\alpha_{s}=\frac{1}{3}$ in which case \cref{eq:eo} gives
\begin{equation}
  \bigg(\frac{E}{\bar{\omega}}\bigg)_{\textrm{LHC}}\simeq 23.
\end{equation}
We note that \cref{eq:rhic} is very close to the estimate given in Ref.~\cite{Frankfurt} some time ago.

\subsection{A comment}
In this section we shall indicate why we do not believe that there is a simple rule for evaluating the spectrum $\omega dI^{\textrm{Nucleus}-\textrm{Proton}}/d\omega$ in two-jet production in terms of the external charges of the partons. We begin with the process gluon $\rightarrow$ quark-antiquark where a simple rule was found in Ref.~\cite{Arleo}. Let us try to see how that rule comes about in our calculation. For simplicity we suppose the transverse coordinate of the emitted gluon $z_{\perp}$ lies in the interval $E/\omega M^{2}<z^{2}<(E/\omega M)^{2}$. Referring to \cref{eq:initialabcone,eq:finalcom}, and taking $Q_{s}^{2}x^{2}$ small as well as factoring out $Q_{s}^{2}x^{2}/8$, the spectrum both for initial and for final state radiation is
\begin{equation}
  \omega\frac{d I^{\textrm{Initial}}}{d\omega}=\omega\frac{d I^{\textrm{Final}}}{d\omega}=\frac{\alpha_{s}N_{c}}{\pi}\int^{(E/\omega M)^{2}}_{E/\omega M^{2}}\frac{dz^{2}}{z^{2}}
\end{equation}
which looks just like bremsstrahlung from a gluon. The interference term given in \cref{eq:crossfullone,eq:crossone} is
\begin{equation}
  \omega\frac{dI^{\textrm{Interference}}}{d\omega}=-\frac{\alpha_{s}N_{c}}{\pi}\int^{(E/\omega M)^{2}}_{E/\omega M^{2}}\frac{dz^{2}}{z^{2}}e^{-Q_{s}^{2}z^{2}/4}.
\end{equation}
Thus one-half of the initial plus final state radiation is canceled if $Q_{s}E/\omega M\ll 1$. The radiation which is emitted when $Q_{s}^{2}z^{2}/4<1$ is due to the hard collision, not due to the multiple scattering. The key to why exactly $\frac{1}{2}$ of the initial and final state radiation is canceled by interference lies in \cref{eq:crossfullone,eq:qqexpand}. Without radiation the four scattering terms in \cref{eq:qqexpand} would be given by \cref{eq:norad}, the factor defining the rate of scattering without radiation. The factor $Q_{s}^{2}x^{2}/16$ on the right-hand side of \cref{eq:qqexpand} is a factor of 2 smaller accounting for the fact that only $\frac{1}{2}$ of the initial and final state radiation is canceled by the interference term.

Now to the main point. The factor of 2 difference between the $Q_{s}^{2}x^{2}/8$ on the right-hand side of \cref{eq:norad} and the $Q_{s}^{2}x^{2}/16$ on the right-hand side of \cref{eq:qqexpand} is not universal. In the process quark $\rightarrow$ quark-gluon the production rate is given by the factor
\begin{equation}
  1+1-2S(x)S(x/2)\simeq \frac{5Q_{s}^{2}x^{2}}{16}
\end{equation}
while the interference term has the scattering factor given by \cref{eq:acbbb}
\begin{equation}
  S^{2}(z)+S^{2}(z-x/2)-2S(x)S(z)S(z-x/2)\simeq S^{2}(z)\frac{Q_{s}^{2}x^{2}}{4}.
\end{equation}
So now the scattering factor is reduced by a factor of $\frac{4}{5}$ in the interference term for radiation. This is the $\frac{4}{5}$ on the far right in \cref{eq:qgregionone}. We do not understand why the radiation factor is $\frac{4}{5}$ here and $\frac{1}{2}$ in gluon $\rightarrow$ quark-antiquark. Since it is the interference term alone which shows up in $\omega dI^{\textrm{Nucleus}-\textrm{Proton}}/d\omega$ we are unable to see a general pattern for the radiation factors which come in for energy loss differences between nuclear and proton targets. As a final comment we note that in quark $\rightarrow$ quark-gluon final state radiation and initial radiation are not the same [compare the $\frac{1}{2}$ on the far right of \cref{eq:qgfullabctwo} with the $\frac{13}{10}$ on the far right of \cref{eq:qgtotalabc}] in contrast to the case of gluon $\rightarrow$ quark-antiquark.

\section{Energy loss in heavy quarkonium production in proton-nucleus collisions: The color evaporation model\label{sec:quarkonium}}
Now let us switch the topic to the calculation of the energy loss in heavy quarkonium production in proton-nucleus collisions. We still view the process in a frame where a gluon from the proton splits into a heavy quark-antiquark pair, each of which has a mass $M\gg Q_{s}$, scattering on a nucleus. The transverse separation of the quark and antiquark is roughly $1/M$ and the transverse momentum of the (anti)quark is much less than $M$. Here we shall focus on two different major models for the heavy quarkonium production: the color evaporation model \cite{Fritzsch,Halzen,Gluck,Barger} and the color singlet model \cite{Kharzeev,Dominguez}. In the color evaporation model the heavy quark-antiquark pair stays in a color octet state and is neutralized by some soft gluon emissions during the subsequent hadronization process. These gluons are different from the one that leads to the energy loss. In the color singlet model the transition from a octet state to a singlet state happens via the last inelastic scattering with the nucleus. This color transition mechanism has been discussed in detail in Refs.~\cite{Kharzeev,Dominguez}.

In the color evaporation model comparing with the gluon $\rightarrow$ quark-antiquark jet production, one sees that the quark mass $M$ plays the role of a hard scale, which is the transverse momentum of the jets in the two-jet case. Moreover, in the two-jet calculation we do not have to take the color transition into account, which is also the case for a quarkonium production in the evaporation model. Therefore the calculation of heavy quarkonium production in the color evaporation model and two-jet calculation are very similar, so we shall focus on the evaporation model first and then study the color singlet model in detail in the next section. 

In the color evaporation model, almost all the calculations we have done for the gluon $\rightarrow$ quark-antiquark jet production can be directly used for the heavy quarkonium production. The major difference is that now the phase space of the gluon radiation is more restricted. Due to the heavy mass of the quark-antiquark pair the small-angle final state gluon radiation is forbidden. The transverse momentum $k_{\perp}$ of a radiated gluon should be greater than $\omega M/E$, below which radiation is not possible. This suppression of small-angle radiation from a heavy mass object is known as the ``dead cone'' phenomenon \cite{Dokshitzer}. The allowed gluon radiation phase space of the heavy quark-antiquark pair coincides with that of the two-jet production when the gluon is emitted coherently from the two jets. That is the gluon radiation calculation in heavy quarkonium production is the same as the large-angle gluon radiation part in the two-jet case. Furthermore, due to the suppression of small-angle gluon emission, we no longer have gluon radiation inside the opening angle between the quark and antiquark, i.e. the first term in \cref{eq:finalcom} should not be included in the present calculation while the rest of the calculation follows exactly the same as the two-jet case. Therefore, the gluon radiation spectrum is still given by \cref{eq:radspec,eq:radspecqs}, with $M$ now interpreted as the mass of the heavy quark. The phenomenology for this case has been carried out in Refs.~\cite{Peigne,Arleo}.

\section{Energy loss in heavy quarkonium production in proton-nucleus collisions: The color singlet model\label{sec:colorsinglet}}
In the color singlet model the color octet quark-antiquark pair becomes a color singlet via either the last inelastic interaction, a mechanism that we will discuss in \cref{sec:singletno}, or a gluon emission that carries a certain amount of energy from the system. However, the transition of the color state only happens once; multiple transitions are suppressed in the large $N_{c}$ limit. In contrast to the color evaporation model the following calculation cannot be applied to proton-proton collisions, simply because in the color singlet model more than one gluon exchange is required to produce the heavy quarkonium. So in the following calculation we only calculate the $J/\psi$ production cross section with energy loss in proton-nucleus collisions without attempting to compare it with proton-proton collisions. Another important aspect of the color singlet model calculation is that we shall take the $J/\psi$ production wave function into account. We will see later the $J/\psi$ wave function puts a quite strong constraint on the form of the $S$ matrix. The cross section reads
\begin{equation}
  \frac{d\sigma_{gA\rightarrow J/\psi X}}{d^{2}b}=\int^{1}_{0}dz\int\frac{d^{2}x_{\perp}}{4\pi}\Phi(x_{\perp},z)\int^{1}_{0}dz'\int \frac{d^{2}x'_{\perp}}{4\pi}\Phi^{*}(x'_{\perp},z')\int \frac{d\omega}{\omega}\,\mathcal{S}(x_{\perp},x'_{\perp};\omega).
  \label{eq:cocs}
\end{equation}
where $\mathcal{S}(x_{\perp},x'_{\perp};\omega)$ is the scattering factor, which will be replaced by, for example, \cref{eq:ssctt}, of the quark-antiquark-gluon system propagating in the presence of a nucleus and $\Phi(x_{\perp},z)=\psi_{J/\psi}^{*}\psi_{g}$ is the $J/\psi$ production wave function which takes the following form \cite{Dominguez,Kowalski,Soyez,Watt}:
\begin{equation}
  \label{eq:wavefunction}
  \Phi(x_{\perp},z)=\frac{g}{\pi\sqrt{2N_{c}}}\Big\{M^{2}K_{0}(Mx_{\perp})\phi_{T}(x_{\perp},z)-\big[z^{2}+(1-z)^{2}\big]MK_{1}(Mx_{\perp})\partial_{x_{\perp}}\phi_{T}(x_{\perp},z)\Big\}
\end{equation}
with
\begin{equation}
  \phi_{T}(x_{\perp},z)=N_{T}z(1-z)\exp\bigg(-\frac{x_{\perp}^{2}}{2R_{T}^{2}}\bigg)
\end{equation}
and where $N_{T}=1.23$, $R_{T}^{2}=6.5\, \textrm{GeV}^{-2}$ \cite{Soyez}. Note that the wave function \cref{eq:wavefunction} contains modified Bessel functions, $K_{0}(Mx_{\perp})$ and $K_{1}(Mx_{\perp})$, which are even functions of $x_{\perp}$. Hence odd terms in either $x_{\perp}$ or $x'_{\perp}$ cannot survive in the calculation of $\mathcal{S}(x_{\perp},x'_{\perp};\omega)$. In the following calculation we will immediately neglect terms that are odd in $x_{\perp}$ or $x'_{\perp}$, which leads to great simplification.

Moreover, we still classify the graphs in the same way as shown in \cref{fig:twojets}. However, the reader should keep in mind that the quark-antiquark pair in \cref{fig:twojets} should be in a color singlet in the final state in the following calculation, hence we put a superscript ``$s$'' on the letter indicating the class. The color restriction, together with the property of the $J/\psi$ wave function, tremendously reduces the number of graphs we have to calculate. Nevertheless the calculation of the last inelastic scattering, the $\xi$ integration [see \cref{eq:singletsmatrix}], becomes complicated for the initial-state-final-state interference terms. The allowed phase space of the gluon radiation is still the same as that in the color evaporation model.

Instead of going directly into the energy loss calculation we shall first review the color singlet model that has been discussed in Ref.~\cite{Dominguez}. This is a good starting place to let the reader see the difference between the calculation in a color singlet model and that of the two-jet and heavy onium production in the color evaporation model. Additional complexities come because we have to integrate over the places that the color transition happens, which makes the evaluation of the $S$ matrix completely different. So let us start with evaluating the $S$ matrix of a quark-antiquark propagating through a nucleus without energy loss while a color transition of the quark-antiquark pair happens during the scattering process.

\subsection{Quarkonium production without energy loss\label{sec:singletno}}
\begin{figure}[h]
  \centering
  \includegraphics[width=13cm]{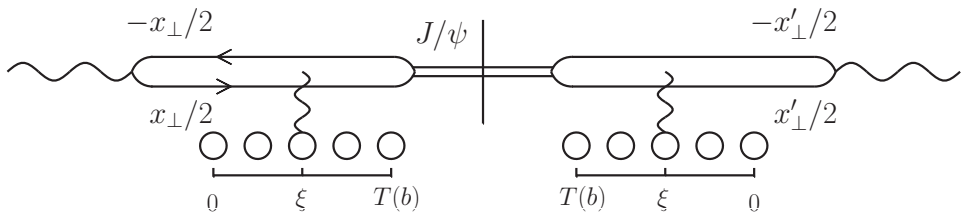}
  \caption{A heavy quark-antiquark pair propagates through a nucleus without energy loss. The circles denote nucleons in the nucleus. The last inelastic scattering happens at a longitudinal position $\xi$. Interactions before or after the last inelastic interaction are not shown.}
  \label{fig:singlet}
\end{figure}
Let us first calculate the $S$ matrix of a heavy quarkonium converting from a color octet state to a color singlet state while propagating through a nucleus as shown in \cref{fig:singlet} \cite{Kharzeev,Dominguez}. Since we do not directly measure the quark-antiquark pair, we distinguish the transverse coordinates of the quark-antiquark pair in the amplitude and that in the complex conjugate amplitude. The quark-antiquark pair is converted to a color singlet state via the last inelastic scattering in the nucleus, which is indicated by a gluon line from a nucleon in \cref{fig:singlet}. The inelastic scattering happening at a longitudinal coordinate $\xi$ brings in a scattering factor
\begin{equation}
  \label{eq:sfactor}
  \frac{Q_{s}^{2}}{4T(b)}x\cdot x'
\end{equation}
where $T(b)$ is the length of the nuclear matter at an impact parameter $b$. Before the last inelastic scattering the quark-antiquark pair is in an octet state, so elastic scatterings can occur off a single quark or antiquark in the amplitude or complex conjugate amplitude, or inelastic scatterings involving the quark (antiquark) in both the amplitude and complex conjugate amplitude. All these possible interactions are not drawn in \cref{fig:singlet}. Then all the scatterings that happen before the last inelastic scatterings give
\begin{equation}
  \label{eq:before}
  \exp\bigg[-\frac{(x-x')^{2}Q_{s}^{2}}{16}\frac{\xi}{T(b)}\bigg].
\end{equation}
After the last inelastic scattering the quark-antiquark pair is in a color singlet state, then only elastic scattering can happen, which gives another factor
\begin{equation}
  \label{eq:after}
  \exp\bigg[-\frac{1}{8}(x^{2}+x'^{2})Q_{s}^{2}\bigg(1-\frac{\xi}{T(b)}\bigg)\bigg].
\end{equation}
Putting \cref{eq:sfactor,eq:before,eq:after} together and integrating over all the possible places that the last inelastic scattering can happen we find \cite{Dominguez}
\begin{align}
  S(x,x')&=\int^{T(b)}_{0}\!d\xi\ \frac{x\cdot x'}{4T(b)}Q_{s}^{2}\exp\bigg[-\frac{(x-x')^{2}Q_{s}^{2}}{16}\frac{\xi}{T(b)}-\frac{(x^{2}+x'^{2})Q_{s}^{2}}{8}\bigg(1-\frac{\xi}{T(b)}\bigg)\bigg]\nonumber\\
  &=\frac{4x\cdot x'}{(x+x')^{2}}\big[e^{-\frac{Q_{s}^{2}}{16}(x-x')^{2}}-e^{-\frac{Q_{s}^{2}}{8}(x^{2}+x'^{2})}\big].
  \label{eq:sno}
\end{align}
When we add one additional gluon to the above process, the calculation can become quite complicated, because the gluon radiation can also become another way of implementing the color transition. Furthermore in the color singlet model we have to distinguish two different kinematic regions depending on the value of the saturation momentum: $Q_{s}^{2}x_{\perp}^{2}\ll 1$ and $Q_{s}^{2}x_{\perp}^{2}\sim 1$ with $x_{\perp}^{2}$ always the size of $1/M^{2}$. Thus we analyze these two different cases separately in the following.

\subsection{Energy loss in region $Q_{s}^{2}x_{\perp}^{2}\ll 1$}
In this region $Q_{s}^{2}\ll M^{2}$, terms that are linear in $Q_{s}^{2}$, i.e. $Q_{s}^{2}(x\cdot x')^{2}dz^{2}/z^{4}$ terms, vanish. We shall look for logarithmic terms like $Q_{s}^{4}(x\cdot x')^{2}dz^{2}/z^{2}$, which are roughly the size of $(Q_{s}/M)^{4}$.

\subsubsection{Purely final state radiation\label{sec:singletf}}
With the restrictions that the quark-antiquark pair must be a color singlet and has even parity in the final state one can immediately see that the graphs in classes $A$ and $B$, illustrated in \cref{fig:twojets}, give no contribution. The only final state emission contribution comes from 
\begin{equation}
  C^{(s)}_{33}=\frac{\alpha_{s}N_{c}}{2\pi^{2}}\int d^{2}z\bigg[\frac{z-x/2}{(z-x/2)^{2}}-\frac{z+x/2}{(z+x/2)^{2}}\bigg]\cdot \bigg[\frac{z-x'/2}{(z-x'/2)^{2}}-\frac{z+x'/2}{(z+x'/2)^{2}}\bigg]S^{2}\big((x'-x)/2\big).
  \label{eq:scthreet}
\end{equation}
The $d^{2}z$ integration in \cref{eq:scthreet} is not logarithmic so that there is no logarithmic contribution from the purely final state radiation.

\subsubsection{Purely initial state radiation\label{sec:singleti}}
Again, graphs in classes $A$ and $B$ give no contribution. The nonzero contributions come from $C^{(s)}_{11}$ and $C^{(s)}_{22}$. For $C^{(s)}_{11}$ the color conversion can happen via an inelastic scattering in the nucleus. We have, for the $C_{11}^{(s)}$ contribution to $\mathcal{S}(x_{\perp},x'_{\perp};\omega)$ in \cref{eq:cocs},
\begin{equation}
  C^{(s)}_{11}=\frac{\alpha_{s}N_{c}}{\pi^{2}}\int \frac{d^{2}z}{z^{2}}S(x,x')
  \label{eq:scoo}
\end{equation}
where $S(x,x')$ is given by \cref{eq:sno}. Since we are in the region where $Q_{s}^{2}x^{2}\ll 1$ we can simply expand the exponents in $S(x,x')$ and only keep the leading term that is even in $x_{\perp}$ and $x'_{\perp}$ due to the parity of the $J/\psi$ wave function. That is 
\begin{equation}
  \label{eq:snoexpand}
  S(x,x')\simeq \frac{Q_{s}^{4}}{64}(x\cdot x')^{2}.
\end{equation}
\Cref{eq:scoo} becomes
\begin{equation}
  C^{(s)}_{11}=\frac{\alpha_{s}N_{c}}{\pi}\frac{Q_{s}^{4}}{64}(x\cdot x')^{2}\int^{1/\mu^{2}}_{E/\omega M^{2}} \frac{dz^{2}}{z^{2}}.
  \label{eq:sscoo}
\end{equation}
For $C^{(s)}_{22}$ the quark-antiquark pair remains in a color octet state after the gluon radiation and it is again the nucleus that converts the pair to a color singlet. Otherwise if the gluon radiation converts the quark-antiquark pair to a singlet and then the pair scatters purely elastically off the nucleus, one can easily see that the quark-antiquark pair has odd parity in $x_{\perp}$ and $x'_{\perp}$. Thus we arrive at
\begin{equation}
  \label{eq:sctt}
  C^{(s)}_{22}=\frac{\alpha_{s}N_{c}}{2\pi^{2}}\int d^{2}z\bigg[\frac{(z-x/2)\cdot (z-x'/2)}{(z-x/2)^{2}(z-x'/2)^{2}}+\frac{(z+x/2)\cdot (z+x'/2)}{(z+x/2)^{2}(z+x'/2)^{2}}\bigg]S(x,x').
\end{equation}
The terms in the brackets can be expanded as
\begin{equation}
  \label{eq:expandpart}
  \bigg[\frac{(z-x/2)\cdot (z-x'/2)}{(z-x/2)^{2}(z-x'/2)^{2}}+\frac{(z+x/2)\cdot (z+x'/2)}{(z+x/2)^{2}(z+x'/2)^{2}}\bigg]\simeq \frac{1}{z^{2}}\bigg(1+\frac{1}{4}\frac{x\cdot x'}{z^{2}}\bigg).
\end{equation}
Using \cref{eq:snoexpand,eq:expandpart} and keeping the leading logarithmic term, we write \cref{eq:sctt} as
\begin{equation}
  C^{(s)}_{22}=\frac{\alpha_{s}N_{c}}{\pi}\frac{Q_{s}^{4}}{64}(x\cdot x')^{2}\int^{E/\omega M^{2}}_{1/M^{2}}\frac{d z^{2}}{z^{2}}.
  \label{eq:onessctt}
\end{equation}

\subsubsection{Initial-state-final-state interference terms}
\begin{figure}[h]
  \centering
  \includegraphics[width=12cm]{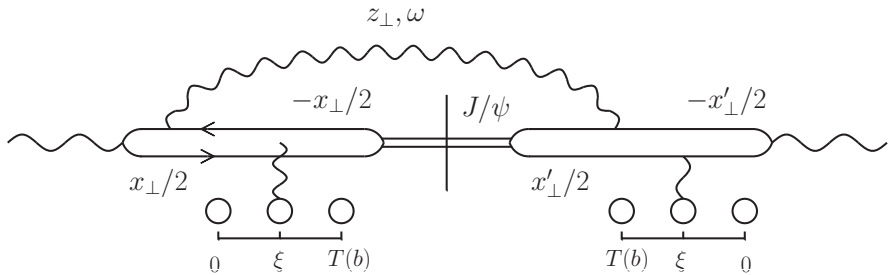}
  \caption{One typical diagram coming from $C^{(s)}_{23}$ in the color singlet model.}
  \label{fig:singletrad}
\end{figure}

Here the nonzero contributions come from $C^{(s)}_{13}=C_{31}^{(s)}$ and $C^{(s)}_{23}=C^{(s)}_{32}$. As we will see later the $S$ matrices describing the scattering for $C^{(s)}_{13}$ and $C^{(s)}_{23}$ are the same with the only difference between them being the gluon emission amplitude. Now the evaluation of the $S$ matrix is slightly more complicated than what we have done in \cref{sec:singletf,sec:singleti}, because in the previous calculation the way that the color transition happens is the same in both the amplitude and the complex conjugate amplitude, either via a gluon radiation or the last inelastic scattering from the nucleus. However, here we have a mixture of the two ways of making the color conversion, i.e. if the quark-antiquark pair is converted to a color singlet via the last inelastic scattering in the nucleus in the amplitude (complex conjugate amplitude) then the gluon radiation must convert the octet state to a singlet state in the complex conjugate amplitude (amplitude). In order to have a clear understanding of the calculation let us take one typical diagram from $C^{(s)}_{23}$ and study it in detail. Furthermore, as we have mentioned before the only difference between $C_{13}^{(s)}$ and $C_{23}^{(s)}$ is the gluon emission part with the $S$ matrix being the same for all of them. Thus let us focus on \cref{fig:singletrad} and evaluate its $S$ matrix part first.
Again, in \cref{fig:singletrad}, the gluons from the circles, which now represent the nucleons, denote the longitudinal places, $\xi$, where the last inelastic scatterings happen. The gluon line attached to the center of the quark-antiquark pair in the amplitude indicates that the gluon can be connected to either the quark or antiquark, while in the complex conjugate amplitude the gluon can only have one possible connection, as indicated in \cref{fig:singletrad}. The other connection of the gluon line is suppressed in large $N_{c}$. The scattering factor that the last inelastic scattering introduces then becomes
\begin{equation}
  \frac{Q_{s}^{2}}{4T(b)}x\cdot (x'/2-z)
\end{equation}
which is different from \cref{eq:sfactor}. Note that now in the amplitude the color transition is made by the last inelastic scattering in the nucleus while in the complex conjugate amplitude the transition to a color singlet state is made by the gluon radiation. Thus the inelastic scattering shown in the complex conjugate amplitude in \cref{fig:singletrad} is not necessary for the process to occur and that interaction could be moved to the amplitude with the scattering occurring off the radiated gluon, which is not shown in \cref{fig:singletrad}. This way of counting the ``inelastic scattering'' factor eventually gives $x\cdot(x'/2-z)$ instead of $x\cdot x'/2$ in \cref{eq:sfactor}. Then we can write down the $S$ matrix
\begin{align}
  S^{(s)}(x,x',z)&=\int^{T(b)}_{0}d\xi\, \exp\bigg[-\frac{Q_{s}^{2}}{8}(z+x/2)^{2}\frac{\xi}{T(b)}-\frac{Q_{s}^{2}}{32}(x-x')^{2}\frac{\xi}{T(b)}-\frac{Q_{s}^{2}}{8}(z+x'/2)^{2}\frac{\xi}{T(b)}\bigg]\nonumber\\
  &\phantom{=}\times \exp\bigg[-\frac{Q_{s}^{2}}{8}x^{2}\frac{T(b)-\xi}{T(b)}-\frac{Q_{s}^{2}}{8}(z+x'/2)^{2}\frac{T(b)-\xi}{T(b)}-\frac{Q_{s}^{2}}{8}(z-x'/2)^{2}\frac{T(b)-\xi}{T(b)}\bigg]\nonumber\\
  &\phantom{=}\times\frac{Q_{s}^{2}}{4T(b)}x\cdot (x'/2-z).
  \label{eq:singletsmatrix}
\end{align}
The first exponential in \cref{eq:singletsmatrix} comes from the interactions before the last inelastic scattering. The quark-antiquark pair is in the color octet state and interactions can happen elastically and inelastically. The second exponential comes from the interactions after the last inelastic scattering. In the amplitude the quark-antiquark pair is already converted to a color singlet via the last inelastic scattering so further interactions on the quark-antiquark pair should be purely elastic, so only an $x^{2}$ term appears and no $(z-x/2)^{2}$ and $(z+x/2)^{2}$ terms. While, on the contrary, in the complex conjugate amplitude, the quark-antiquark is still in the color octet state and color conversion happens via the later gluon radiation. Thus in the complex conjugate amplitude inelastic scatterings are still allowed after the inelastic scattering at $\xi$, which leads to the $(z+x'/2)^{2}$ and $(z-x'/2)^{2}$ terms in the second exponential while no $x'^{2}$ term appears. Finishing the $\xi$ integration in \cref{eq:singletsmatrix} we find
\begin{equation}
  S^{(s)}(x,x',z)=\frac{1}{b-a}\big(e^{-\frac{Q_{s}^{2}}{8}a}-e^{-\frac{Q_{s}^{2}}{8}b}\big)
  \label{eq:fullsinglets}
\end{equation}
where $a=2z^{2}+\frac{1}{2}x'^{2}+x^{2}$ and $b-a=(z-\frac{1}{2}x)\cdot (x+x')$. The gluon emitted from the $(-x'_{\perp}/2)$ line in the complex conjugate amplitude can be moved to the $(x'_{\perp}/2)$ line without introducing any additional factor of $N_{c}$ because the quark-antiquark is in a color singlet state after the gluon emission. Including this additional graph we finally have
\begin{equation}
  2C^{(s)}_{23}=-2\cdot \frac{\alpha_{s}N_{c}}{2\pi^{2}}\int d^{2}z\frac{z+x/2}{(z+x/2)^{2}}\cdot \bigg[\frac{z+x'/2}{(z+x'/2)^{2}}-\frac{z-x'/2}{(z-x'/2)^{2}}\bigg]S^{(s)}(x,x',z)+
  \begin{pmatrix}
    x\rightarrow -x\\
    x'\rightarrow -x'
  \end{pmatrix}
  \label{eq:scts}
\end{equation}
where the second term, i.e. $(x\rightarrow -x,\, x'\rightarrow -x')$, takes into account the graphs that the gluon is emitted from the $(x_{\perp}/2)$ line in the amplitude. Using
\begin{equation}
  \frac{z+x/2}{(z+x/2)^{2}}\cdot \bigg[\frac{z+x'/2}{(z+x'/2)^{2}}-\frac{z-x'/2}{(z-x'/2)^{2}}\bigg]\simeq \frac{x\cdot x'}{(z^{2})^{2}}\bigg[z^{2}+\frac{1}{2}x\cdot x'-z\cdot (x+x')\bigg]
\end{equation}
and
\begin{equation}
  S^{(s)}(x,x',z)=-\frac{Q_{s}^{4}}{128}x\cdot (x'-2z)\bigg[4z^{2}+z\cdot (x+x')-\frac{1}{2}x\cdot x'\bigg],
  \label{eq:expandsinglets}
\end{equation}
\cref{eq:scts} can be simplified to
\begin{equation}
  2C^{(s)}_{23}=\frac{\alpha_{s}N_{c}}{\pi}\frac{Q_{s}^{4}}{128}(x\cdot x')^{2}\int_{1/M^{2}}^{E/\omega M^{2}}\frac{dz^{2}}{z^{2}}.
  \label{eq:tsctt}
\end{equation}
The nucleus, in the amplitude in \cref{fig:singletrad}, cannot distinguish whether the gluon radiation is from the initial gluon or the quark-antiquark pair, so the $S$ matrix for $C^{(s)}_{13}$ is still given by \cref{eq:fullsinglets}. Thus we have
\begin{equation}
  2C^{(s)}_{13}=-2\cdot \frac{\alpha_{s}N_{c}}{2\pi^{2}}\int d^{2}z\frac{2z}{z^{2}}\cdot \bigg[\frac{z-x'/2}{(z-x'/2)^{2}}-\frac{z+x'/2}{(z+x'/2)^{2}}\bigg]S^{(s)}(x,x',z)+
  \begin{pmatrix}
    x\rightarrow -x\\
    x'\rightarrow -x'
  \end{pmatrix}.
  \label{eq:scot}
\end{equation}
Using
\begin{equation}
  \frac{z}{z^{2}}\cdot \bigg[\frac{z-x'/2}{(z-x'/2)^{2}}-\frac{z+x'/2}{(z+x'/2)^{2}}\bigg]\simeq \frac{z\cdot x'}{(z^{2})^{2}}
\end{equation}
and \cref{eq:expandsinglets}, we arrive at
\begin{equation}
  2C^{(s)}_{13}=-\frac{\alpha_{s}N_{c}}{\pi}\frac{Q_{s}^{4}}{64}(x\cdot x')^{2}\int^{(E/\omega M)^{2}}_{E/\omega M^{2}} \frac{dz^{2}}{z^{2}}.
  \label{eq:tscot}
\end{equation}
\Cref{eq:tsctt,eq:tscot} have been written  assuming $(E/\omega M)^{2}<1/Q_{s}^{2}$. If this is not the case the limits in \cref{eq:tsctt,eq:tscot} should be changed to ensure $z_{\perp}^{2}<1/Q_{s}^{2}$.  For the moment let us suppose that $(E/\omega M)^{2}<1/Q_{s}^{2}$ then adding \cref{eq:onessctt,eq:tsctt,eq:tscot} gives
\begin{equation}
  C^{(s)}_{22}+2C^{(s)}_{23}+2C^{(s)}_{13}=\frac{\alpha_{s}N_{c}}{\pi}\frac{Q_{s}^{4}}{128}(x\cdot x')^{2}\ln\frac{E}{\omega}
  \label{eq:sqqctotal}
\end{equation}
while if $E/\omega M^{2}<1/Q_{s}^{2}<(E/\omega M)^{2}$ one has
\begin{equation}
  C_{22}^{(s)}+2C_{23}^{(s)}+2C_{13}^{(s)}=\frac{\alpha_{s}N_{c}}{\pi}\frac{Q_{s}^{4}}{128}(x\cdot x')^{2}\bigg[3\ln \frac{E}{\omega}-2\ln\frac{M^{2}\omega}{E Q_{s}^{2}}\bigg].
  \label{eq:sqqctotaltwo}
\end{equation}
We recall that the initial state radiation given in \cref{eq:sscoo} is already included in the parton distribution for the gluon initiating process. 

Let us estimate energy loss using \cref{eq:sqqctotal,eq:sqqctotaltwo}. Dividing these equations by \cref{eq:snoexpand} gives the energy spectrum. We use the criterion \cref{eq:avg} to determine $\bar{\omega}$. Assuming $E/\bar{\omega}M^{2}<1/Q_{s}^{2}$, but $(E/\bar{\omega}M)^{2}>1/Q_{s}^{2}$ we get
\begin{equation}
  \label{eq:radspeccolorsingletone}
  \frac{1}{2}=\int^{E}_{\bar{\omega}}\frac{dI}{d\omega}d\omega=\frac{\alpha_{s}N_{c}}{2\pi}\int^{E}_{EQ_{s}/M}\frac{d\omega}{\omega}\ln\frac{E}{\omega}+\frac{\alpha_{s}N_{c}}{2\pi}\int^{EQ/M}_{\bar{\omega}}\frac{d\omega}{\omega}\bigg[3\ln\frac{E}{\omega}-2\ln\frac{M^{2}\omega}{EQ_{s}^{2}}\bigg].
\end{equation}
This leads to
\begin{equation}
  \frac{5}{2}\ln^{2}\frac{E}{\bar{\omega}}-4\ln\frac{M}{Q_{s}}\ln\frac{E}{\bar{\omega}}=\frac{\pi}{\alpha_{s}N_{c}}-2\ln^{2}\frac{M}{Q_{s}}.
  \label{eq:radspecsingletest}
\end{equation}
For example, for $\ln(M/Q_{s})=1$ and $\alpha_{s}=\frac{1}{3}$, \cref{eq:radspecsingletest} gives $E/\bar{\omega}\simeq 6$.

\subsection{Energy loss in region $Q_{s}^{2}x_{\perp}^{2}\simeq Q_{s}^{2}/M^{2}\sim 1$}
In this region we shall also look for a logarithmic contribution in $d^{2}z_{\perp}/z_{\perp}^{2}$. In this region the calculation is rather simple and we can directly use the results we have obtained in the previous section. Note that now the transverse coordinate of the radiated gluon satisfies $z_{\perp}^{2}>1/M^{2}\sim x_{\perp}^{2}\sim 1/Q_{s}^{2}$, which makes the $S$ matrix containing the factor $e^{-Q_{s}^{2}z^{2}}$  small. We see that only the initial-state-final-state interference terms have $z_{\perp}$ dependence in the $S$ matrices [see \cref{eq:scts,eq:scot}], so they are suppressed in this region. The only nonzero contributions come from the initial and final state gluon radiation. However, from \cref{eq:scthreet} we can see that $C^{(s)}_{33}$, the only term allowed by the parity of the $J/\psi$ wave function in the final state radiation, is not logarithmic in $z_{\perp}$. While for the purely initial state radiation $C^{(s)}_{11}$ should be considered as part of the gluon wave function of the proton instead of the energy loss, so we are left only with $C^{(s)}_{22}$. Since $Q_{s}^{2}x^{2}_{\perp}\sim 1$ in this region we can no longer expand the exponents in the $S$ matrix, but we can still expand the gluon emission wave function. Using \cref{eq:sno,eq:sctt,eq:expandpart} we find
\begin{equation}
  C_{22}^{(s)}=\frac{\alpha_{s}N_{c}}{\pi}\frac{4x\cdot x'}{(x+x')^{2}}\big[e^{-Q_{s}^{2}(x-x')^{2}/16}-e^{-Q_{s}^{2}(x^{2}+x'^{2})/8}\big]\int^{E/\omega M^{2}}_{1/M^{2}}\frac{dz^{2}}{z^{2}}
  \label{eq:ssctt}
\end{equation}
with $Q_{s}^{2}x_{\perp}^{2}\sim 1$. Dividing \cref{eq:ssctt} by \cref{eq:sno} gives
\begin{equation}
  \label{eq:radspeccolorsinglettwo}
  \omega\frac{dI}{d\omega}=\frac{\alpha N_{c}}{\pi}\ln\frac{E}{\omega}.
\end{equation}
Now we get
\begin{equation}
  \frac{E}{\bar{\omega}}=\exp\bigg(\sqrt{\frac{\pi}{\alpha_{s}N_{c}}}\bigg)
\end{equation}
giving $E/\bar{\omega}\simeq 6$ for $\alpha_{s}=\frac{1}{3}$.
\begin{acknowledgments}
  One of the authors, A.~M., wishes to thank Fran\c{c}ois Arleo and St\'ephane Peign\'e for a very useful discussion at the \'Ecole Polytechnique where they explained the work in Refs.~\cite{Peigne,Arleo}. The authors also wish to thank Jianwei Qiu for some helpful discussions. This work is supported, in part, by the US Department of Energy. As we were completing this work we found out that a similar study by S.~Peign\'e \textit{et al.} \cite{Kolevatov} was also in the final stages of completion. We have had a stimulating and very useful correspondence with F.~ Arleo and S.~ Peign\'e on the subjects covered in our works. Although the technical details are different in our approaches there is much overlap and there is agreement on major issues.
\end{acknowledgments}

\appendix

\section{Appendix\label{app:ps}}
\begin{figure}[h]
  \centering
  \includegraphics[width=12cm]{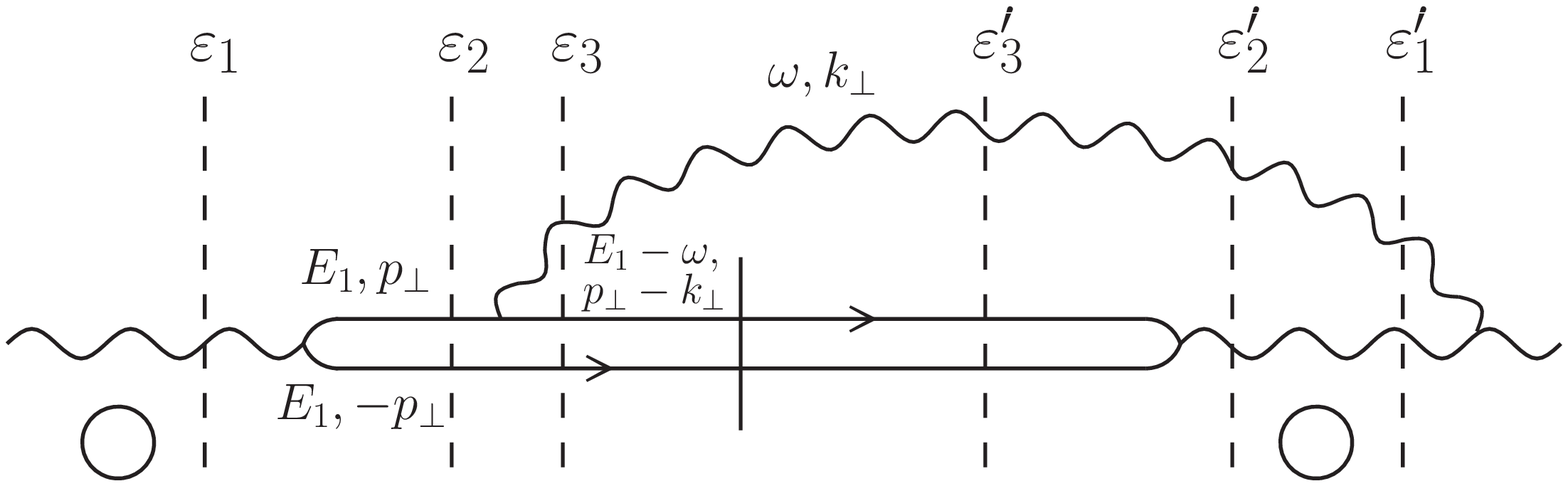}
  \caption{One graph from $A_{31}$. The vertical dashed lines indicate the intermediate states.}
  \label{fig:example}
\end{figure}
In this appendix we shall briefly show how one determines logarithmic regions of integration in the types of graphs considered in this paper. We take the graph $A_{31}$ as an example shown in \cref{fig:example} and indicate the labelings of the lines in terms of momenta where we suppose the quark and antiquark share the longitudinal momentum of the incoming gluon equally. The light cone perturbation theory denominators for the graph are
\begin{equation}
  \textrm{Denominators}=(\varepsilon_{3}-\varepsilon_{2})(\varepsilon_{3}-\varepsilon_{1})(\varepsilon'_{3}-\varepsilon'_{2})(-\varepsilon'_{1}),
\end{equation}
where the incoming gluon has energy $E=2E_{1}$ and no transverse momentum. It is straightforward to find, when $\omega/E_{1}\ll 1$ and $k_{\perp}/p_{\perp}\ll 1$, 
\begin{subequations}
  \begin{align}
  \varepsilon'_{1}&\simeq \frac{k_{\perp}^{2}}{2\omega},\\
  \varepsilon'_{3}-\varepsilon'_{2}&\simeq 2\frac{p_{\perp}^{2}}{2E_{1}}=\frac{2p_{\perp}^{2}}{E},\\
  \varepsilon_{3}-\varepsilon_{2}&\simeq \frac{(k_{\perp}-\omega p_{\perp}/E_{1})^{2}}{2\omega},\\
  \varepsilon_{3}-\varepsilon_{1}&\simeq \frac{k_{\perp}^{2}}{2\omega}+2\frac{p_{\perp}^{2}}{2E_{1}}=\frac{k_{\perp}^{2}}{2\omega}+\frac{2p_{\perp}^{2}}{E}.
\end{align}
\end{subequations}
In addition to the denominators the gluon emission and absorption summed over polarizations give a factor 
\begin{equation}
  k_{\perp}\cdot \bigg(k_{\perp}-\frac{\omega}{E_{1}}p_{\perp}\bigg).
\end{equation}
Thus the graph will have a logarithmic integral, $dk_{\perp}^{2}/k_{\perp}^{2}$, when
\begin{equation}
  \bigg(\frac{\omega}{E_{1}}p_{\perp}\bigg)^{2}\ll k_{\perp}^{2}\ll 4p_{\perp}^{2}\frac{\omega}{E}.
\end{equation}
Using $E=2E_{1}$ and $p_{\perp}^{2}\simeq M^{2}$ one gets
\begin{equation}
  \bigg(\frac{2\omega M}{E}\bigg)^{2}<k_{\perp}^{2}<4\frac{\omega}{E}M^{2}.
  \label{eq:appkbound}
\end{equation}
Changing to coordinate space, $k_{\perp}\leftrightarrow 2/z_{\perp}$, \cref{eq:appkbound} becomes
\begin{equation}
  \frac{E}{\omega M^{2}}<z_{\perp}^{2}<\bigg(\frac{E}{\omega M}\bigg)^{2}.
  \label{eq:appzbound}
\end{equation}
Logarithmic regions for all other graphs can be estimated similarly. The constant factors in \cref{eq:appzbound} are our best ``guess'' as to where limits of the logarithmic integration should be set.

\end{document}